\documentclass[sigconf,natbib,anonymous=false,review=false,screen=true]{acmart}

\AtBeginDocument{%
  \providecommand\BibTeX{{%
    Bib\TeX}}}

\usepackage{booktabs} 
\usepackage{algorithm, algpseudocode}
\usepackage{amsmath}
\usepackage{graphics}
\usepackage{epsfig}
\usepackage{graphicx}
\usepackage{xcolor}
\usepackage{subfigure}
\usepackage{balance}
\usepackage{multirow}
\usepackage{mathrsfs}
\usepackage{acronym}
\usepackage{placeins}
\usepackage{xcolor}
\usepackage[inline]{enumitem}
\usepackage{fancyhdr}
\usepackage{bm}
\usepackage{tabularx}
\usepackage{ragged2e} 
\newcolumntype{L}{>{\RaggedRight\hangafter=1\hangindent=0em}X}
\usepackage{makecell}
\usepackage{graphicx}
\usepackage{soul} 
\usepackage{color} 

\usepackage[skins]{tcolorbox} 

\AtBeginDocument{%
  \providecommand\BibTeX{{%
    \normalfont B\kern-0.5em{\scshape i\kern-0.25em b}\kern-0.8em\TeX}}}

\acrodef{CRS}{conversational recommender system}
\acrodef{KG}{knowledge graph}
\acrodef{CR}{conversational recommendation}
\newcommand{\OurMethod}{ECR}

\makeatletter
\g@addto@macro\normalsize{%
  \abovedisplayskip 1.5pt plus1pt 
  \belowdisplayskip 1.5pt plus1pt
  \abovedisplayshortskip  0pt plus1pt%
  \belowdisplayshortskip  0pt plus1pt
}

\makeatother

\copyrightyear{2024}
\acmYear{2024}
\setcopyright{rightsretained}
\acmConference[RecSys '24]{18th ACM Conference on Recommender Systems}{October 14--18, 2024}{Bari, Italy}
\acmBooktitle{18th ACM Conference on Recommender Systems (RecSys '24), October 14--18, 2024, Bari, Italy}\acmDOI{10.1145/3640457.3688133}
\acmISBN{979-8-4007-0505-2/24/10}



\begin{document}

\title{Towards Empathetic Conversational Recommender Systems}


\author{Xiaoyu Zhang}
\orcid{0000-0002-5667-1036}
\affiliation{
  \institution{Shandong University}
  \city{}
  \country{}
}
\authornote{Work done during an internship at 2023 Tencent Rhino-Bird Research Elite Program.}
\email{tinyoctopus1999@gmail.com}

\author{Ruobing	Xie}
\orcid{0000-0003-3170-5647}
\affiliation{
  \institution{Tencent}
  \city{}
  \country{}
}
\email{xrbsnowing@163.com}

\author{Yougang	Lyu}
\orcid{0009-0000-1082-9267}
\affiliation{
  \institution{Shandong University\\ University of Amsterdam}
  \city{}
  \country{}
}
\email{youganglyu@gmail.com}

\author{Xin Xin}
\orcid{0000-0001-6116-9115}
\affiliation{
  \institution{Shandong University}
  \city{}
  \country{}
}
\email{xinxin@sdu.edu.cn}

\author{Pengjie Ren}
\orcid{0000-0003-2964-6422}
\affiliation{
  \institution{Shandong University}
  \city{}
  \country{}
}
\email{jay.ren@outlook.com}

\author{Mingfei	Liang}
\orcid{0000-0002-0599-1806}
\affiliation{
  \institution{Tencent}
  \city{}
  \country{}
}
\email{aesopliang@tencent.com}

\author{Bo Zhang}
\orcid{0000-0003-2942-1311}
\affiliation{
  \institution{Tencent}
  \city{}
  \country{}
}
\email{nevinzhang@tencent.com}

\author{Zhanhui	Kang}
\orcid{0009-0006-5151-4222}
\affiliation{
  \institution{Tencent}
  \city{}
  \country{}
}
\email{kegokang@tencent.com	}

\author{Maarten	de Rijke}
\orcid{0000-0002-1086-0202}
\affiliation{
  \institution{University of Amsterdam}
  \city{}
  \country{}
}
\email{m.derijke@uva.nl}

\author{Zhaochun Ren}
\orcid{0000-0002-9076-6565}
\affiliation{
  \institution{Leiden University}
  \city{}
  \country{}
}
\authornote{Corresponding author.}
\email{z.ren@liacs.leidenuniv.nl}

\def\authors{Xiaoyu Zhang, Ruobing Xie, Yougang Lyu, Xin Xin, Pengjie Ren, Mingfei Liang, Bo Zhang, Zhanhui Kang, Maarten de Rijke, and Zhaochun Ren}

\renewcommand{\shortauthors}{Xiaoyu Zhang et al.}


\begin{abstract}
\Acfp{CRS} are able to elicit user preferences through multi-turn dialogues. They typically incorporate external knowledge and pre-trained language models to capture the dialogue context. 
Most CRS approaches, trained on benchmark datasets, assume that the standard items and responses in these benchmarks are optimal. However, they overlook that users may express negative emotions with the standard items and may not feel emotionally engaged by the standard responses. This issue leads to a tendency to replicate the logic of recommenders in the dataset instead of aligning with user needs.
{To remedy this misalignment, we introduce \emph{empathy} within a CRS. With empathy we refer to a system's ability to capture and express emotions. We propose an \textbf{e}mpathetic \textbf{c}onversational \textbf{r}ecommender (ECR) framework.}

\OurMethod{} contains two main modules: emotion-aware item recommendation and emotion-aligned response generation. Specifically, we employ user {emotions} to refine user preference modeling for accurate recommendations. To generate human-like emotional responses, \OurMethod{} applies retrieval-augmented prompts to fine-tune a pre-trained language model aligning with emotions and mitigating hallucination. 
{To address the challenge of insufficient supervision labels, we enlarge our empathetic data using emotion labels annotated by large language models and emotional reviews collected from external resources}. 
We propose novel evaluation metrics to capture user satisfaction in real-world CRS scenarios.
Our experiments on the ReDial dataset validate the efficacy of our framework in enhancing recommendation accuracy and improving user satisfaction.

\end{abstract}


\maketitle

\section{Introduction}

Advances in conversational systems have led to the integration of natural language conversations with recommender systems, culminating in the development of \acfp{CRS}~\citep{DBLP:conf/nips/LiKSMCP18}. 
A crucial aspect of CRSs is to elicit user preferences through multi-turn dialogues, with two main subtasks: \emph{item recommendation} and \emph{response generation}~\cite{DBLP:journals/aiopen/GaoLHRC21}. 
A prominent challenge is the lack of sufficient contextual information for accurately modeling user preferences. 
Some research~\citep{DBLP:conf/emnlp/ChenLZDCYT19,DBLP:conf/kdd/ZhouZBZWY20} integrates knowledge graphs (KGs) and models user preferences based on entities from KGs mentioned in the dialogues. 
Recent work~\citep{DBLP:journals/tois/0002ZXLCL23, DBLP:conf/kdd/WangZWZ22, DBLP:conf/naacl/YangHLZY22} centers on using pre-trained language models (PLMs) to enhance the system's understanding of dialogue context.
Despite these advances, existing \ac{CRS} models still do not fully align with user needs. 
These models are trained on conversational recommendation training datasets. 
But the presumption that the standard items and responses in the dataset are optimal leads to a tendency of \ac{CRS} to replicate the logic of recommenders in the dataset instead of addressing user needs.

\smallskip\noindent
\textbf{Using empathy to address misalignment.}
The above misalignment hinders the development of \acp{CRS}. \citet{lerner2015emotion} have proposed that \emph{emotions} are crucial in human decision-making processes. Their work suggests that capturing emotions expressed in user utterances within dialogues is prominent for achieving accurate user preference modeling for item recommendation. 
People tend to favor agents simulating human beings by exhibiting \emph{emotions}~\citep{Cominelli2021PromisesAT}. 
Adopting emotion-rich expressions in response generation can enliven the user experience and contribute to user satisfaction. Serving users in a natural, human-like way by capturing and expressing emotions is a necessary development for CRSs in terms of aligning with user needs, thereby offering benefits for users and providers of recommender systems.
We introduce \emph{empathy} within a \ac{CRS}, defining it as the system's capacity to capture and express emotions~\citep{10.1093/acprof:oso/9780199659180.003.0005}. 
Through empathy, we aim to accurately distinguish and fulfill user needs, both during {item recommendation} and {response generation}. 

\begin{figure}[t]
    \centering
    \includegraphics[width=1.05\linewidth]{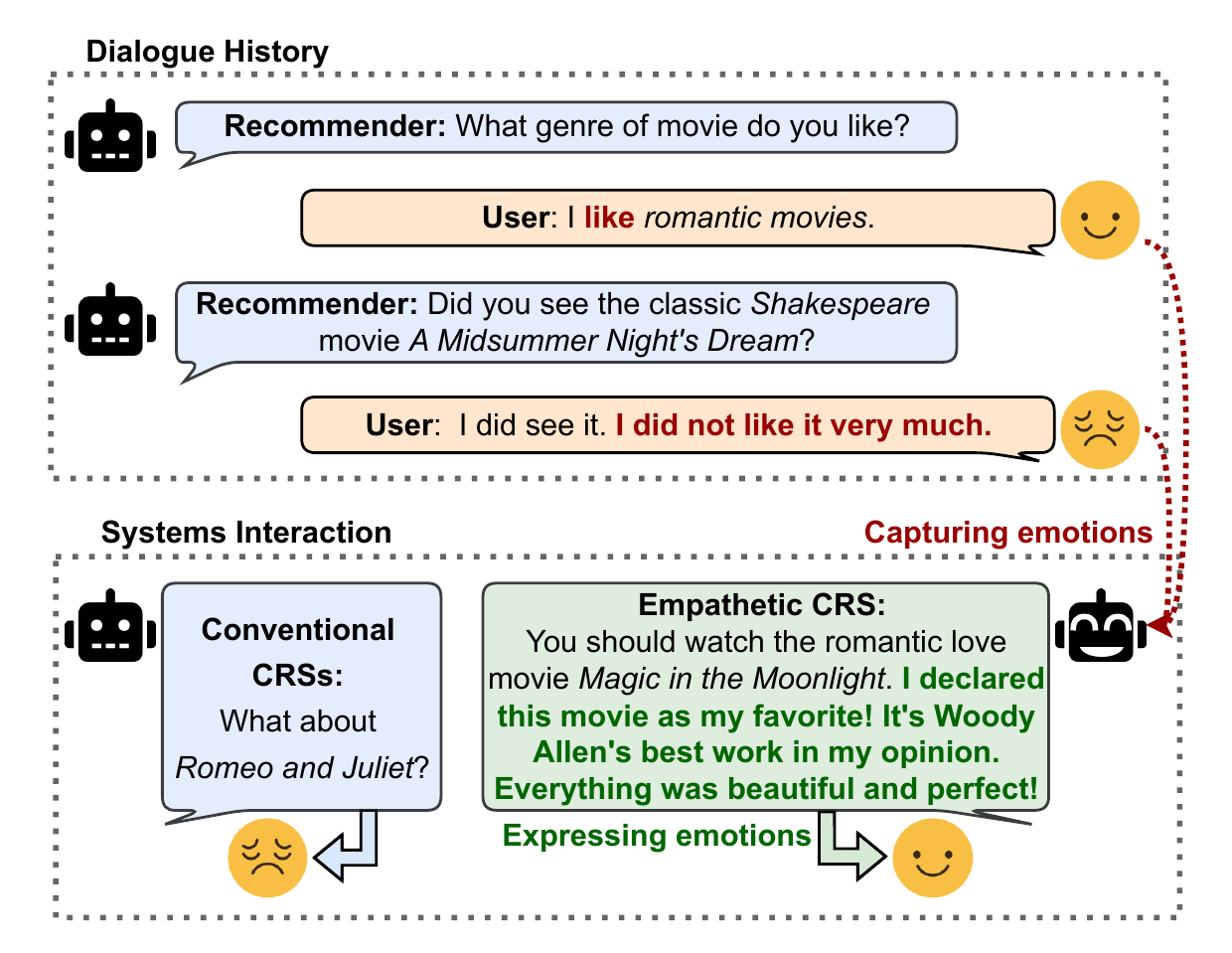}
    \caption{An example of a conversation on movie recommendation between a user and the system. Text conveying user emotions is highlighted in red font. System responses expressing emotions are marked in green font.}
    \label{fig:ECR_example}
\end{figure}

\smallskip
\noindent
\textbf{Integrating empathy into recommendation and generation.}
We analyze the need to integrate empathy into item recommendation and response generation subtasks, respectively. For item recommendation, existing approaches often assume that all entities mentioned in dialogues reflect user preferences and that all items suggested by recommenders meet user expectations.
This hypothesis disregards subtle cues of user {emotions} expressed in natural language for modeling user preferences.
As illustrated in Figure~\ref{fig:ECR_example}, a conventional CRS might infer that the user likes ``Shakespeare'' and ``A Midsummer Night's Dream'' mentioned by the recommender, while overlooking that the user expresses negative emotions towards them during the dialogue. Thus such systems recommend the wrong item ``Romeo and Juliet.''
For response generation, existing methods are trained on the standard responses from datasets, which tend to be short and lack narratives, often resulting in inconsistencies or a lack of emotional engagement.
As shown in Figure~\ref{fig:ECR_example} (bottom), the conventional CRSs' response only contains the item name, which may diminish user satisfaction when interacting with the system. In contrast, based on capturing and expressing emotions, an empathetic CRS recommends a reasonable item with a persuasive response.

\smallskip\noindent
\textbf{Challenges.} To construct empathetic CRSs, we face two major challenges: (i) how to accurately model user preferences using emotions; and 
(ii) how to generate emotional responses contributing to user satisfaction.
To address these challenges, we propose an \textbf{e}mpathetic \textbf{c}onver\-sational \textbf{r}ecom\-mender (ECR) framework comprising two key modules: \emph{emotion-aware item recommendation} and \emph{emotion-aligned response generation}. 
For the emotion-aware item recommendation module, we integrate user {emotions} with entities in the utterance to augment user preference modeling. We also propose a training strategy to minimize the impact of incorrect labels in the dataset. For the emotion-aligned response generation module, we fine-tune a pre-trained language model (PLM) to express emotions. 
To avoid hallucination, we retrieve relevant knowledge from KGs as a part of generation prompts. 
Existing \ac{CRS} datasets lack user emotion labels and emotional responses. 
To enlarge the available empathetic training data, we use large language models (LLMs) to discern nuanced emotions in the dialogue history; then, we collect emotional reviews as an informative external resource for fine-tuning the PLM to generate emotional responses.

Since existing evaluation metrics ignore the impact of emotions, we introduce novel metrics for CRSs, aiming at better reflecting user satisfaction in real-world CRS scenarios.
For item recommendation, we adopt the Area Under the Curve (AUC) metric to assess the model's accuracy in modeling user preferences. 
AUC requires that items receiving positive feedback from users should have a higher possibility of being recommended than those with negative feedback. 
For response generation, we move beyond traditional metrics like BLEU or ROUGE, opting instead to use five subjective metrics: emotional intensity, emotional persuasiveness, logic persuasiveness, informativeness, and lifelikeness. 
Experiments on the ReDial benchmark dataset confirm the effectiveness of our proposed framework.

\smallskip\noindent
\textbf{Contributions.} The contributions of this paper are as follows:
\begin{enumerate*}[label=(\roman*)]
    \item 
    To bridge the gap between system outputs and user needs, we define empathy within a CRS and propose a novel framework \OurMethod{}.
    \item We augment user preference modeling by integrating their {emotions}, with a new training strategy to minimize the impact of incorrect labels. 
    \item We fine-tune a PLM to express emotions and apply retrieval-augmented prompts to mitigate hallucination.
    \item We use LLMs to annotate user emotions and collect emotional reviews from external resources as empathetic CRS training data, which facilitates future research in this area.
    \item We propose new evaluation metrics tailored to user satisfaction in real-world CRS scenarios, and our experimental results demonstrate that \OurMethod{} significantly outperforms baselines on the ReDial dataset.
\end{enumerate*}

\section{Related Works}
The literature on CRSs~\citep{DBLP:conf/nips/LiKSMCP18, DBLP:conf/kdd/ZhouZBZWY20, DBLP:conf/kdd/WangZWZ22} can be classified into \emph{attribute-based} CRSs and \emph{generation-based} CRS~\citep{DBLP:conf/recsys/Zhang23a}. 
Attribute-based CRSs~\citep{DBLP:conf/kdd/Christakopoulou16, DBLP:conf/cikm/ZhouZWWZWW20} predominantly employ fixed response templates and predefined actions for user interaction. The primary objective of most methodologies within this category is to minimize the number of turns required to complete the recommendation task~\citep{DBLP:conf/wsdm/Lei0MWHKC20, DBLP:conf/kdd/LeiZ0MWCC20}. \citet{DBLP:conf/sigir/DengL0DL21} and \citet{DBLP:conf/kdd/LeiZ0MWCC20} use KGs to improve the recommendation performance. However, they still overlook the importance of generating high-quality natural language, which can be detrimental to the overall user experience.

Unlike attribute-based CRSs, generation-based CRSs~\citep{DBLP:conf/nips/LiKSMCP18,DBLP:conf/coling/ZhouZZWW20} focus on making recommendations using free-form text, which creates considerable flexibility to influence how a dialogue continues. \citet{DBLP:conf/nips/LiKSMCP18} use an auto-encoder for recommendation and a hierarchical RNN for response generation. However, a challenge these systems face is the lack of sufficient contextual information for accurately discerning user preferences~\citep{DBLP:conf/cikm/HeXJSLFMKM23}. 
Research indicates that CRSs can be enhanced by incorporating additional sources of knowledge. 
\citet{DBLP:conf/emnlp/ChenLZDCYT19} integrates KGs to enhance the user representation and propose an end-to-end framework. 
\citet{DBLP:conf/kdd/ZhouZBZWY20} incorporate both word-oriented and entity-oriented KGs. 
Through reasoning based on the entities from KGs mentioned in the dialogues~\citep{DBLP:conf/emnlp/MaTH21,DBLP:conf/emnlp/ZhouWHH21, DBLP:journals/corr/abs-2201-02732, DBLP:conf/acl/Li0QMYXC23}, this integration further enhances the logical accuracy of recommendation and response interpretability.
Subsequent research also introduces reviews~\citep{DBLP:conf/acl/LuBSMCWH21, DBLP:journals/corr/abs-2201-02732} and in-text knowledge~\citep{DBLP:conf/sigir/LiXZ0Z022, DBLP:conf/wsdm/Zhang0LLRCMR23, DBLP:conf/sigir/RenTLR0XLRC22} to assist user preference modeling.

Recent work on generation-based CRSs have centered on integrating LLMs into CRSs~\citep{DBLP:journals/tois/0002ZXLCL23, DBLP:conf/kdd/WangZWZ22, DBLP:conf/naacl/YangHLZY22}.  UniCRS~\citep{DBLP:conf/kdd/WangZWZ22} addresses the recommendation and generation subtasks in a unified approach with prompt tuning. 
\citet{DBLP:conf/cikm/HeXJSLFMKM23} conduct an in-depth analysis of LLMs for zero-shot CRS. And \citet{DBLP:conf/emnlp/WangTZWW23} develop an interactive evaluation method using LLM-based user simulators. 
However,~\citet{DBLP:conf/recsys/DaiSZYSXS0X23} demonstrate that traditional collaborative filtering recommendation models, with adequate data, significantly outperform LLMs. 
Moreover, while LLMs are proficient in conversational aspects, they face limitations in conversational recommendation tasks, particularly in capturing user emotional engagement.

Our study aligns with the generation-based CRSs.
A major problem of recent generation-based CRSs is their misalignment with user preferences. We integrate empathy into CRS, prioritizing user needs as our goal. Similarly to our approach, methods for empathetic response generation~\citep{DBLP:conf/aaai/LiLRRC22, DBLP:conf/emnlp/YangRYZCCWSJL23, DBLP:conf/coling/LiCRRTC20} detect and respond to user emotions. These methods are tailored for chat and not easily adapted to CRSs. Some traditional recommender systems have enhanced collaborative filtering by incorporating sentiment analysis~\citep{DBLP:journals/fi/KarabilaDEAM23, DBLP:journals/tcss/KumarDR20,DBLP:journals/sensors/DangGP21}. However, these works only focus on the analysis of item reviews rather than real-time multi-turn natural language dialogues. 

\section{Preliminaries} %
\label{section: Preliminaries Work: UniCRS}
\subsection{Problem Formulation}

\smallskip\noindent
\textbf{Notation.} Given $t-1$ dialogue turns, the dialogue history $D_{t-1}$ consists of a sequence of utterances from both recommenders and users, i.e., $D_{t-1}=\{u_k^r, u_k^u\}^{t-1}_{k=1}$, where each utterance $u_k^* = \{w_j\}^{|u_k^*|}_{j=1}$ is composed of a sequence of words. 
For simplicity, we concatenate all utterances from $D_{t-1}$ into a single word sequence $D = \{w_q\}^{n_w}_{q=1}$, where $n_w$ represents the total number of words in $D_{t-1}$.
To incorporate knowledge about entities mentioned in the dialogue, we set an external knowledge graph (e.g., DBpedia~\cite{DBLP:journals/ws/BizerLKABCH09}) as $\mathcal{G}=(\mathcal{E},\mathcal{L})$, consisting of triples {$\mathcal{T} = \left \langle e_{h}, l, e_{t}\right \rangle$,} where $e_h \in \mathcal{E}$ and $e_{t}\in \mathcal{E}$ are the head and tail entities, {$l \in \mathcal{L}$} reflects the relation between $e_h$ and $e_t$. $\mathcal{E}$ and {$\mathcal{L}$} denote the sets of entities and relations. We define $I$ as the entire set of items, all of which are included in the entities of $\mathcal{G}$, i.e., $I \in \mathcal{E}$. Entities in each utterance $u_k^*$ are identified as $E_k^* = \{e_j\}^{|E_k^*|}_{j=1}$. Each item $i_j$ within $E_k^*$ is linked with user feedback $f_{i_j}$, indicating whether the user likes it.
Similarly, we combined all entities mentioned in $D_{t-1}$ into an entity list $E_l = \{e_q\}^{n_e}_{q=1}$, where $n_e$ is the count of entities in the dialogue history. Here, we refer to the entities mentioned in the dialogue history as \emph{local entities}. Correspondingly, we refer to entities co-occurring with the local entities in the training dataset as \emph{global entities}, which will be detailed in Section~\ref{subsubsection: Emotion-aware item representing}.

\smallskip\noindent
\textbf{Task outputs.} At the $t$-th turn, a CRS  
\begin{enumerate*}[label=(\roman*)]
\item selects a set of target items $I_t = \{{i_k}\}^{|{I_t}|}_{k=1}$ from the entire item set $I$, and 
\item generates a response utterance {$u_t^r$} for the user.
\end{enumerate*}

\subsection{Backbone Framework}
Since UniCRS~\cite{DBLP:conf/kdd/WangZWZ22} unifies the recommendation and generation into a prompt learning paradigm by PLM, i.e., DialoGPT~\citep{DBLP:conf/acl/ZhangSGCBGGLD20}, which is the state-of-the-art method in using PLMs, we adopt it as our backbone framework. It encompasses three primary modules:
\begin{enumerate}[leftmargin=*]
    \item \textbf{Semantic fusion module}:  
    Initially, UniCRS fuses the semantic spaces of dialogues and KGs for knowledge alignment and enrichment. It obtains a word embedding matrix 
    and a local entity embedding matrix. 
    Then it associates two kinds of embedding matrices via a bilinear transformation, yielding the fused word representations $\tilde{\bm{W}} =  \left[\bm{\tilde{w}_1}\right.$; \ldots; $\left.\bm{\tilde{w}_{n_w}}\right]$, and the fused local entity representations $\tilde{\bm{E_l}} = \left[\bm{\tilde{e}_1}; \ldots; \bm{\tilde{e}_{n_e}}\right]$. 

    \item \textbf{Response generation module}: UniCRS prompts a PLM to generate the response $u^r_t$, which is designated as ``\emph{recommendation response.}'' The prompt for this module consists of the fused word representations $\tilde{\bm{W}}$, generation task-specific soft tokens $\bm{S_{gen}}$, and the dialogue history $D$:
    \begin{equation}
        \label{conv_prompt}
        \begin{aligned}
            C_{gen}^{r} = \left[\tilde{\bm{W}}; \bm{S_{gen}}; D\right].
        \end{aligned}
    \end{equation}
    Note that UniCRS replaces all items appearing in the recommendation response with a special token [MASK], which is later filled following the item recommendation subtask.

    \item \textbf{Item recommendation module}: Given $u_t^r$ from the response generation subtask, the recommendation prompt consists of the fused local entity representations $\tilde{\bm{E_l}}$, recommendation task-specific soft tokens $\bm{S_{rec}}$, the dialogue history $D$, and $u_t^r$: 
    \begin{equation}
        \label{rec_prompt}
        \begin{aligned}
            C_{rec} = \left[\tilde{\bm{E}_l}; \bm{S_{rec}}; D; u_t^r\right].
        \end{aligned}
    \end{equation}
    {The response generation and item recommendation modules both use cross-entropy loss for prediction.}
\end{enumerate}

\noindent%
Although UniCRS shows promise in using PLMs, its optimization still relies on standard answers provided by datasets and ignores user emotions, which limits its ability to track user needs.
It inspires our subsequent endeavors in instantiating \OurMethod{} based on UniCRS. 
Note that our proposed framework can extend beyond UniCRS and be seamlessly adapted to other CRSs with modifications.
\section{Method}
\label{sec:method}
\begin{figure*}
    \centering
    \includegraphics[width=1.0\linewidth]{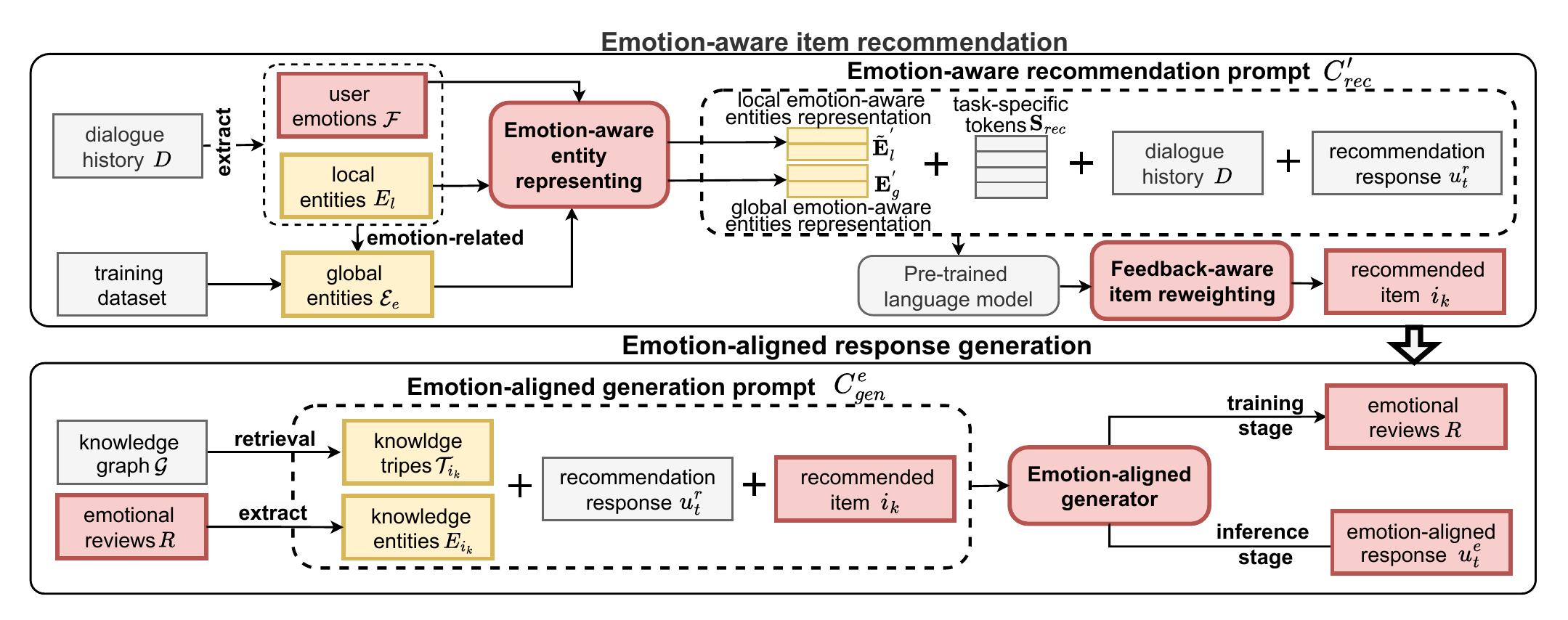}
    \caption{Overview of \OurMethod{}. \OurMethod{} has two key modules: (i) emotion-aware item recommendation for better user preference understanding, and (ii) emotion-aligned response generation for engaging conversations.}
    \Description{The framework of \OurMethod{}.}
    \label{fig:ECR_moel}
\end{figure*}

In this section, we introduce our empathetic data enlargement process (Section~\ref{subsection: Data enlargement}) and two key modules of \OurMethod{}: emotion-aware item recommendation (Section~\ref{subsec: Reflect aware Item Recommendation}) and emotion-aligned response generation (Section~\ref{subsec: Empathetic Response Generation}). Figure~\ref{fig:ECR_moel} shows an overview of \OurMethod{}.

\subsection{Empathetic Data Enlargement}
\label{subsection: Data enlargement}
\subsubsection{User Emotion Extraction}
\label{subsubsection: User emotion extraction}
Existing datasets lack explicit supervisory signals for identifying {user emotions}. 
To address the problem, we employ GPT-3.5-turbo~\cite{chatgpt} to initially annotate user emotions in 5,082 utterances from the ReDial dataset. 
We limit the number of annotated emotions per utterance to a maximum of two labels. In annotating with GPT-3.5-turbo~\citep{chatgpt} for utterance-level user emotions, we adopted nine emotion labels: ``like,'' ``curious,'' ``happy,'' ``grateful,'' ``negative,'' ``neutral,'' ``nostalgia,'' ``agreement,'' and ``surprise.'' 
The ``negative'' label, denoting adverse emotions, accounted for 8.0\%. See Appendix ~\ref{app: User emotion extraction} for a more detailed annotation process.
Based on the annotations, we fine-tune a GPT-2 model, which achieves 87.75\% in terms of Recall@2 in categorizing emotions. 
We applied this model to annotate emotions for each user utterance $u_k^u$ in the ReDial dataset and set a threshold $\beta$ to retain relevant emotion label $f$. 
For each utterance $u_k^u$, we obtain its utterance-level user emotions $\mathcal{F}_{u_k^u} = \{f_j\}_{j=1}^{|\mathcal{F}_{u_k^u}|}$, along with the probabilities associated with each emotional label, denoted as $\mathcal{P}_{u_k^u} = \{p_j\}_{j=1}^{|\mathcal{P}_{u_k^u}|}$, where $|\mathcal{F}_{u_k^u}| = |\mathcal{P}_{u_k^u}|$.

\subsubsection{Emotional Response Construction.}
\label{subsubsection: review data construction}
There is an abundance of reviews about consumer experiences on the web.
These reviews are frequently imbued with the writers' personal experiences and emotions, with comprehensive information about various facets of the items being reviewed. Hence, emotional reviews serve as optimal resources for emotional responses.
Considering the emphasis of our study is on recommendation task and positive emotions contribute more to effectively persuasive and engaging interactions~\citep{ KWORTNIK2007324}, we only adopted top-rated (10/10) reviews rich in positive emotions, rather than expressing different types of emotions according to user emotions.
To construct a emotional review database $R$, we collect movie reviews from IMDb.\footnote{\url{https://www.imdb.com/}}
Each emotional review is retrieved according to the item $i_r$ in the training dataset. The constructed emotional review database $R = \{r_j\}^{|R|}_{j=1}$ comprises a sequence of emotional review sentences. Each emotional review sentence $r = \{w_j\}^{|r|}_{j=1}$ consists of a sequence of words. From each emotional review sentence $r$, we extract a list of entities $E_r = \{e_j\}^{|E_r|}_{j=1}, e_j \in \mathcal{E}$. We then retrieve a set of knowledge triples $\mathcal{T}_r = \{\langle i_r, l, e_{j}\rangle, e_{j}\in E_r\}^{|\mathcal{T}_r|}_{j=1}$ from the knowledge graph $\mathcal{G}$, where the item $i_r$ is the head entity, the entity $e_j \in E_r$ is the tail entity. 

\subsection{Emotion-aware Item Recommendation}
\label{subsec: Reflect aware Item Recommendation}

\subsubsection{Emotion-aware Entity Representation}
\label{subsubsection: Emotion-aware item representing}
Entities are essential for reflecting user preferences. 
Thus we aim to model the effect of user emotions on the entities. 
Since local entities $E_l$ only reflect user interest exhibited in the ongoing dialogue $D_{t-1}$, which is insufficient for comprehensively exploring user preferences, we collect global entities from the training dataset filtered by user emotions, which encompass the collaborative knowledge shared by all users. 
In general, we model the effect of user emotions on entities both in the dialogue history and in the training data, respectively.

\smallskip\noindent
\textbf{Local emotion-aware entity representing.} To model the effect of user emotions on local entities, we characterize utterance-level user emotions $\mathcal{F}_{u_k^u} = \{f_j\}_{j=1}^{|\mathcal{F}_{u_k^u}|}$ as reflecting emotions towards entities mentioned by the user in the current utterance $E_k^u$ and by the recommender in the preceding one $E_k^r$. Consequently, each local entity $e_j \in E_l$ is linked to an utterance-level user emotions, represented as $\mathcal{F}_{e_j} = \{f_i\}_{i=1}^{|\mathcal{F}_{e_j}|}$ along with corresponding probabilities $\mathcal{P}_{e_j} = \{p_i\}_{i=1}^{|\mathcal{P}_{e_j}|}$.
Hence, we calculate the user emotion representation of each local entity $e_j$ as:
\begin{equation}
\label{emotion_rep}
\begin{aligned}
    &\bm{\mathcal{F}_{e_j}} = \sum^{|\mathcal{F}_{e_j}|}_{i = 1}p_i*v(f_i), 
\end{aligned}
\end{equation}
where $v(f_i)$ denotes the learnable representation of the $i$-th emotion label in the utterance-level user emotions. 
Then, we fuse the user emotion representation $\bm{\mathcal{F}_{e_j}}$ with the local entity representation of $e_j$ to get a local emotion-aware entity representation $\bm{\tilde{e}_j'}$, as follows:
\begin{equation}
\label{fusion}
   \bm{\tilde{e}_j'} =\left[\bm{\tilde{e}_j}; \bm{\mathcal{F}_{e_j}}\right]\mathbf{W}_{\boldsymbol{t}}^{\top} + \boldsymbol{b},
\end{equation}
where $[\cdot;\cdot]$ denotes vector concatenation; $\mathbf{W}_{\boldsymbol{t}}$ and $\boldsymbol{b}$ are learnable parameters aimed at projecting the dimension of the concatenated representation back to the dimension of $\bm{\tilde{e}_j}$. 
We stack all local emotion-aware entity representations into a matrix, denoted as $\bm{\tilde{E_l}'} = \left[\bm{\tilde{e}_1'}; \ldots; \bm{\tilde{e}_{n_e}'}\right]$.

\smallskip\noindent
\textbf{Global emotion-aware entity representing.} We first use utter\-ance-level user emotions to filter global entities and then aggregate their representations. 
Concretely, we assume that if a user exhibits similar emotions towards both $e_j$ and $e_i$ in a conversation, then $e_i$ is globally \emph{emotion-related} to $e_j$. 
Similarly, we define a local entity $e_j$ to be emotion-related to a set of global entities $\mathcal{E}_{e_j} = \{e_{i}\}^{|\mathcal{E}_{e_j}|}_{i=1}$, where $e_j$ and $e_i$ overlap in the most probable $n_f$ emotion labels in $\mathcal{F}_{e_j}$ and $\mathcal{F}_{e_i}$ during their co-occurrence in a conversation. 
Additionally, we calculate the co-occurrence probability of each local entity $e_j$ with its emotion-related global entity $e_i$ from the training dataset, denoted as $P(e_i|e_j)$.
Then, we aggregate the representation of all global entities emotion-related to the local entity $e_j$ as a global entity representation $\bm{\mathcal{E}_{e_j}}$:
\begin{equation}
\label{co_rep}
\begin{aligned}
    &\bm{\mathcal{E}_{e_j}} = \sum^{|\mathcal{E}_{e_j}|}_{i = 1}\bm{e_i}*P(e_i|e_j),
\end{aligned}
\end{equation}
where $\bm{e_i}$ denotes the representation of $e_i$ obtained from the RGCN \citep{DBLP:conf/esws/SchlichtkrullKB18}. Following the Eq.~\ref{fusion}, we calculate the global emotion-aware entity representation $\bm{\mathcal{E}_{e_j}'}$ by integrating $\bm{\mathcal{E}_{e_j}}$ and $\bm{\mathcal{F}_{e_j}}$. 
Finally, we stack global emotion-aware entity representation for each local entity $e_j \in E_l$
into a matrix, denoted as $\bm{E_{g}'} =  \left[\bm{\mathcal{E}_{e_1}'}; \ldots; \bm{\mathcal{E}_{e_{n_e}}'}\right]$.

\smallskip\noindent
\textbf{Emotion-aware recommendation prompt.} To comprehensively model user preferences with their emotions, we use the local emo\-tion-aware entity representation matrix $\bm{\tilde{E}_{l}'}$ and global emotion-aware entity representation matrix $\bm{E_{g}'}$ to update the prompt in Eq.~\ref{rec_prompt}. So we formulate an {emotion-aware recommendation prompt} as:
\begin{equation}
\label{rec_prompt_1}
    C_{rec}' = 
    \left[
    \bm{\tilde{E}_{l}'}; \bm{E_{g}'}; \bm{S_{rec}}; D; u_t^r
    \right].
\end{equation}

\subsubsection{Feedback-aware Item Reweighting}
\label{subsubsection: Emotion-aware item reweighting}
In the preceding section, we employ utterance-level user emotions to track user entity-based preferences. 
In this section, we develop a reweighting strategy that is aware of user feedback ${f}_{i_k}$ on each recommended item $i_k$, aligning with the supervision labels provided in the dataset but ignored by most CRS methods.
Specifically, we introduce a mapping function $m(f_{i_k})$ that converts each user feedback $f_{i_k}$ as a weight scalar. 
The mapping function converts negative or unclear feedback into a lower weight. 
Based on the weight scalars, we rewrite the cross-entropy loss  for item recommendation subtask as: 
\begin{equation}
\label{loss}
    L_{rec}=-\sum_{k =i}^N m(f_{i_k})\log Pr(i_k\mid C_{rec}'),
\end{equation}
where $N$ represents the total number of training instances, and $Pr(i_k\mid C_{rec}')$ refers to the predicted probability of the recommended item $i_k$ given the {emotion-aware recommendation} prompt $C_{rec}'$.

\subsection{Emotion-aligned Response Generation}
\label{subsec: Empathetic Response Generation}
To support an engaging user experience, we generate an \emph{emotion-aligned response} $u_t^{e}$ to enrich the recommendation response $u_t^{r}$ generated by the UniCRS. 
In this section, we construct an emotion-aligned generation prompt and train an emotion-aligned generator.

\subsubsection{Emotion-aligned Generation Prompt}
\label{subsubsection: Emotion-aligned generation prompt}
While PLMs can memorize information from their training corpus, \citet{DBLP:journals/csur/JiLFYSXIBMF23} have shown that PLMs often exhibit hallucinations, which may diminish users' satisfaction with their usage. 
Teaching PLMs to accurately retain knowledge is resource-intensive and challenging. 
Thus, we construct an emotion-aligned generation prompt based on retrieved knowledge to enhance the informativeness in the responses while mitigating hallucination.

Specifically, during the training stage, given the extracted knowledge entities $E_r$ and the retrieved knowledge triples $\mathcal{T}_r$, we transform the entities and triples into word sequences, represented as $S_{\mathcal{T}_r}$ and $S_{E_r}$. 
The prompt for generating emotional review $r$ consists of the word sequence of the knowledge entities $S_{E_r}$, knowledge triples $S_{\mathcal{T}_r}$, and the item name $S_{i_r}$. 
Then, we incorporate the recommendation response $u_t^r$ into the prompt, guiding the model to generate contextually relevant responses. 
The emotion-aligned generation prompt is formally denoted as:
\begin{equation}
\label{conv_2_prompt}
    C_{gen}^{e} = \left[ S_{E_r}; S_{\mathcal{T}}; S_{i_r}; u_t^r\right].
\end{equation}
During the inference stage for generating emotion-aligned responses $u_t^{e}$, we followed the same prompt design as in Eq.~\ref{conv_2_prompt}. 
We retrieve knowledge triples $\mathcal{T}_{i_k} = \{\langle i_k, l, e_{j}\rangle\}^{|\mathcal{T}_{i_k}|}_{j=1}$ from the KG $\mathcal{G}$ using the predicted item $i_k$ as the head entity. 
And we collect a list of knowledge entities $E_{i_k} = \{e_j\}^{|E_{i_k}|}_{j=1}$ that is mentioned at least twice in the reviews corresponding to $i_k$. 
Then, we filter $pn_t$ triples from $\mathcal{T}_{i_k}$ and $pn_e$ entities from $E_{i_k}^{r}$ as a part of the emotion-aligned generation prompt. We simplify the filtering process by random selection, leaving more complex approaches to be explored in the future.

\subsubsection{Emotion-aligned Generator}
To align the model with the persuasive dialogue style and emotions, we fine-tune a PLM  as an emotion-aligned generator to generate the emotion-aligned responses $u_t^{e}$ using the constructed emotional review databases $R$. Specifically, based on the emotional reviews $r$, we employ cross-entropy for training the emotion-aligned generator, as follows:
\begin{equation}
\label{conv_2_loss}
    L_{gen}^{e}(r) =
    -\sum_{j=1}^{|r|}\log\Pr(w_{j}\mid C_{gen}^{e}; w_{<j}),
\end{equation}
where $\Pr(w_{j}\mid C_{gen}^{e}; w_{<j})$ denotes the predicted probability of the word $w_j$ given the prompt $C_{gen}^{e}$ and the words proceeding the $j$-th position. Ultimately, we combine the emotion-aligned response $u_t^{e}$ with the recommendation response $u_t^r$ to formulate the final response delivered to the users.

Following~\citet{DBLP:conf/kdd/WangZWZ22}, we choose DialoGPT~\citep{DBLP:conf/acl/ZhangSGCBGGLD20} as the PLM for emotion-aligned response generation (\OurMethod{}[DialoGPT]). Given DialoGPT's limited parameters, which inherently restrict its linguistic capabilities, we introduce an alternative version by using Llama 2-7B-Chat \cite{DBLP:journals/corr/abs-2307-09288} {to which we refer as \OurMethod{}[Llama 2-Chat]}. 
This choice is motivated by that Llama 2-Chat is an open-source, powerful LLM instruction-tuned on chat tasks. 
This variation allows us to evaluate our framework's performance based on LLMs. 

\section{Experiments}
\label{sec:experiments}
We address the following research questions:
\begin{enumerate*}[label=(\textbf{RQ\arabic*}),leftmargin=*,nosep]
\item Does \OurMethod{} learn user preferences by capturing their emotions to improve the accuracy of recommendation?
\item Is \OurMethod{} capable of expressing emotions in response generation, thereby improving the user satisfaction?
\item How does each component of \OurMethod{} contribute to its overall performance?
\end{enumerate*}

\subsection{Dataset}
\label{subsec:datasets}
The ReDial dataset~\citep{DBLP:conf/nips/LiKSMCP18} is a large-scale CRS dataset, carefully curated by human workers~\citep{DBLP:journals/aiopen/GaoLHRC21}. Consequently, it effectively reflects real-world CR scenarios and fully validates the effectiveness of our method. Considering the significant cost of emotion annotations and evaluations in the generation subtask, we use the ReDial dataset for experiments and plan to extend ECR to other datasets in future work. The ReDial dataset is composed of two-party dialogues between a user and a recommender in the movie domain. 
It contains 10,006 conversations consisting of 182,150 utterances related to 51,699 movies.  
The user feedback towards items recommended in the dataset includes three categories: ``like,'' ``dislike,'' and ``not say.'' 
Previous works~\citep{DBLP:conf/kdd/WangZWZ22,DBLP:journals/corr/abs-2201-02732,DBLP:conf/emnlp/MaTH21} simply treat all the recommended items as positive labels. However, according to \citet{DBLP:conf/nips/LiKSMCP18}, the ``dislike'' and ``not say'' labels are distributed separately at 4.9\% and 14\%, indicating the previous works introduce a large number of incorrect item labels. In contrast, we distinguish between those items with different user feedback. For emotional responses construction, we filter 34,953 reviews related to 4,092 movies for DialoGPT, and 2,459 reviews related to 1,553 movies for Llama 2-7B-Chat. The filtering process is detailed in Appendix ~\ref{appendix: Review preprocessing}.
Following \citep{DBLP:conf/emnlp/ChenLZDCYT19}, we extract entities mentioned in each utterance and review from DBpedia. 

\subsection{Baselines}
\label{subsec:baselines}
For the item recommendation subtask, we compare our method with several CRS approaches: \textbf{KBRD} \citep{DBLP:conf/emnlp/ChenLZDCYT19}, \textbf{KGSF}\citep{DBLP:conf/kdd/ZhouZBZWY20}, \textbf{RevCore}\citep{DBLP:conf/emnlp/MaTH21}, \textbf{UCCR}\citep{DBLP:journals/corr/abs-2201-02732} and \textbf{UniCRS}~\citep{DBLP:conf/kdd/WangZWZ22} to evaluate the effectiveness of \OurMethod{}. Specifically, KBRD first uses KGs to enhance the semantics fusion in recommendation and generation systems. It uses transformer~\citep{DBLP:conf/nips/VaswaniSPUJGKP17} for response generation with enhanced modeling of word weights. KGSF integrates both word-oriented and entity-oriented KGs to refine user representations and employs the transformer for response generation. RevCore introduces a review-enhanced framework, using item reviews for improving recommendations and response generation, with a focus on sentiment-aware review selection. UCCR focuses on comprehensive user modeling by considering multi-aspect information from current and historical dialogues, as well as data from look-alike users. UniCRS unifies the recommendation and generation into a prompt learning paradigm by PLM.

For the response generation subtask, our comparison involves the state-of-the-art CRS model \textbf{UniCRS}~\citep{DBLP:conf/kdd/WangZWZ22}, the powerful open-source dialogue LLM \textbf{Llama 2-7B-Chat}~\cite{DBLP:journals/corr/abs-2307-09288} and two advanced OpenAI models: \textbf{GPT-3.5-turbo-instruct} and \textbf{GPT-3.5-turbo}~\citep{chatgpt}. To make the output deterministic, we set temperature${}=0$ when calling the API. These LLMs are prompted to chat with users, aiming to recommend the item predicted by the recommendation module of \OurMethod{}. They are all provided with the dialogue history for consistency in evaluation.

\subsection{Emotion-enhanced Evaluation Metrics}
\label{subsection:Emotion-enhanced evaluation metrics}

Our evaluation encompasses subjective and objective metrics to assess recommendation and generation performance respectively, which considers the user satisfaction in real-world CRS scenarios. We discuss more details in Appendix ~\ref{Discussion of evaluation metrics}.

\smallskip\noindent
\textbf{Objective evaluation metrics.}
For recommendation evaluation, we employed Recall@$n$ (R@$n$, where $n= 1, 10, 50$) to verify if the top-$n$ recommended items include the target item suggested by the dataset's recommenders. To validate the model's effectiveness in estimating user preferences while negating the logged errors in the dataset, we calculate Recall\_True@$n$ (RT@$n$, where $n= 1, 10, 50$). This metric refines Recall@$n$ but only considers the items that get the user feedback of ``like'' as the standard answers. Additionally, we incorporate the Area Under the Curve (AUC) metric, which emphasizes the ranking order between recommended items linked to the users' positive and negative feedback. 

\smallskip\noindent
\textbf{Subjective evaluation metrics.}
The generation quality is evaluated across five dimensions: emotional intensity (Emo Int), emotional persuasiveness (Emo Pers), logic persuasiveness (Log Pers), informativeness (Info), and lifelikeness (Life). (a) \emph{Emotional intensity} measures the strength of emotions conveyed to users. (b) \emph{Emotional persuasiveness} gauges the capacity to connect with the user emotionally to persuade users. (c) \emph{Logic persuasiveness} evaluates the use of logical reasoning and coherent arguments to persuade users. (d) \emph{Informativeness} determines the utility of useful information provided by the system. (e) \emph{Lifelikeness}  assesses how vivid and engaging the responses are, reflecting their resemblance to natural human communication. 
The scoring range for these metrics is 0 to 9. 

Following \citet{DBLP:conf/emnlp/WangTZWW23}, we employ an LLM-based scorer capable of automatically assigning scores based on specific prompts to alleviate the evaluation reliance on human annotations and randomly sampling 1,000 examples for evaluation. In this context, GPT-4-turbo from the OpenAI serves as the scoring tool. Given the inherent instability in LLMs, we invite three human annotators to assess the reliability of our LLM-based scorer's evaluation results. The annotators are enlisted to rate 200 examples. Additionally, to ensure the robustness of our evaluation, GPT-4 is also employed as an auxiliary scorer, with results detailed in Appendix ~\ref{Discussion of evaluation metrics}.

\subsection{Experimental Settings}
\label{subsec:implementation details}
We implement \OurMethod{} with PyTorch. The embedding size of the emotion label is 48. 
The threshold $\beta$ and $n_f$ are set to 0.1 and 3. 
The amount of knowledge triples $pn_t$ and entity $pn_e$ in the emotion-aligned generation prompt is set to 2 and 4. For the feedback-aware item reweighting strategy, we assign the weight scalar of user feedback ``like,'' ``dislike,'' and ``not say'' to 2.0, 1.0, and 0.5, respectively. The analysis for the hyperparameters, i.e., the weight scalar of user feedback and the amount of knowledge used in the emotion-aligned generation prompt, can be found in Appendix ~\ref{app: Hyperparameters analysis}.
In the emotion-aligned response generation process, we use AdamW~\citep{DBLP:conf/iclr/LoshchilovH19} to optimize the tunable parameters of DialoGPT and fine-tune Llama2-chat with LoRA~\citep{DBLP:conf/iclr/HuSWALWWC22}. 
We set the learning rate for DialoGPT and Llama2-chat to $1e-4$ and $5e-5$, respectively. The batch size is set to 128 for the emotion-aware item recommendation and 16 for the emotion-aligned response generation. The prompts used for the LLM-based scorers and baselines are detailed in Appendix ~\ref{app: Prompts}.

\begin{table}[htbp]
\small
\caption{Objective evaluation of item recommendation. The boldface indicates the best result. Significant improvements over best baseline results are marked with * (t-test, $p \textless 0.05)$.}
\label{tab:recommendation results}
\resizebox{\linewidth}{!}{
\begin{tabular}{l ccccccc}
\toprule
Model & AUC & RT@1 & RT@10 & RT@50 & R@1 & R@10 & R@50\\
\midrule
KBRD & 0.503   & 0.040    &0.182    &0.381    & 0.037 & 0.175 & 0.360 \\ 
KGSF & 0.513  &0.043    & 0.195   & 0.383   & 0.040 &  0.182 & 0.361 \\
RevCore &0.514    & 0.054   & 0.230   & 0.410   & 0.046 & 0.209 & 0.390 \\
UCCR & 0.499  & 0.038  & 0.208   & 0.423   & 0.039 & 0.198 &  0.407  \\
UniCRS & 0.506 &0.052  & 0.229 & 0.439  & 0.047 & 0.212 & 0.414 \\
\midrule 
\OurMethod{}  & \textbf{0.541}\rlap{$^\ast$}  & \textbf{0.055} & \textbf{0.238}\rlap{$^\ast$}  &\textbf{0.452}\rlap{$^\ast$}  & \textbf{0.049} & \textbf{0.220}\rlap{$^\ast$}  &\textbf{0.428}\rlap{$^\ast$}\\
\bottomrule
\end{tabular}}
\end{table}

\begin{table}[htbp]
\small
\caption{Subjective evaluation of LLM-based scorer (GPT-4-turbo) and human annotators for response generation. 
The boldface indicates the best result. Significant improvements over best baseline results are marked with * (t-test, $p \textless 0.05)$.}
\label{tab:generation result}
\resizebox{\linewidth}{!}{
\begin{tabular}{l l ccccc}
\toprule
& Model &Emo Int& Emo Pers &Log Pers& Info & Life\\
\midrule
\multirow{6}{*}{\rotatebox[origin=c]{90}{\textbf{LLM-based scorer}}} 
& UniCRS &0.400 &0.942 &0.793 &0.673 &2.241\\
& GPT-3.5-turbo-instruct &1.706 &3.043 &3.474 &2.975 &4.182\\
& GPT-3.5-turbo &2.215 &3.754 &4.782 &4.147 &5.338\\
& Llama 2-7B-Chat &3.934 &6.030 &5.886 &5.904 &7.129\\
\cmidrule(r){2-7}
& \OurMethod{}[DialoGPT] &4.011 &4.878 &4.736 &5.094 &5.906\\			
& \OurMethod{}[Llama 2-Chat] &\textbf{6.826}\rlap{$^\ast$} &\textbf{7.724}\rlap{$^\ast$} &\textbf{6.702}\rlap{$^\ast$} &\textbf{7.653}\rlap{$^\ast$} &\textbf{8.063}\rlap{$^\ast$}\\
\bottomrule
\multirow{6}{*}{\rotatebox[origin=c]{90}{\textbf{Human annotator}}}
& UniCRS &0.947 &0.775 &1.158 &0.380 &1.805\\
& GPT-3.5-turbo-instruct &2.048 &2.555 &3.265 &1.822 &3.648\\
& GPT-3.5-turbo &2.890 &3.678 &5.323 &3.233 &5.125\\
& Llama 2-7B-Chat &4.432 &6.152 &6.393 &5.713 &7.463\\
\cmidrule(r){2-7}
& \OurMethod{}[DialoGPT] &5.097 & 4.817 &5.398 &4.628  &6.385\\			
& \OurMethod{}[Llama 2-Chat] &\textbf{7.130}\rlap{$^\ast$} &\textbf{7.575}\rlap{$^\ast$} &\textbf{7.403}\rlap{$^\ast$} &\textbf{7.172}\rlap{$^\ast$} &\textbf{8.468}\rlap{$^\ast$}\\

\bottomrule
\end{tabular}}
\end{table}

\subsection{Evaluation on Item Recommendation (RQ1)}

\noindent 
We address RQ1 by evaluating the performance of item recommendation; see Table~\ref{tab:recommendation results}. 
KGSF and RevCore, introducing external knowledge in CRSs, have demonstrated superior performance compared to KBRD, underscoring the significance of external knowledge in recommendations. 
UCCR also performs well on RT@50 and R@50 by extracting user-centric data from cross-session interactions. 
UniCRS, which integrates PLM into CRSs, exhibits the best performance among all baselines on RT@$n$ and R@$n$.
Regarding the AUC, a metric previously overlooked but essential for evaluating a model's full alignment with users' needs, we find that all baselines exhibit poor performance, with AUC values approaching $0.5$. This finding highlights the considerable challenge faced by CRSs in distinguishing between items receiving positive and negative feedback.

We observe that \OurMethod{} significantly outperforms all the baselines. 
Specifically, it shows an improvement of 3.9\% and 3.0\% over UniCRS in RT@10 and RT@50, respectively.
Additionally, \OurMethod{} demonstrates a significant lead in the AUC metric, as indicated by a  6.9\% improvement over UniCRS. These findings confirm the importance of capturing user emotions in enhancing CRSs ability to accurately estimate user preferences and effectively mitigate the impact of incorrect item labels in the dataset.

\subsection{Evaluation on Response Generation (RQ2)}
\label{subsection:Evaluation on Conversation Task}

To analyze whether \OurMethod{} is capable of expressing emotions for better user satisfaction, we conduct a comparison of response generation, evaluated by both LLM-based scorer (GPT-4-Turbo) and human annotators, as shown in Table~\ref{tab:generation result}. 
We observe that the evaluation results from the LLM-based scorer and human annotators are essentially consistent. In comparison to all baseline models, we discovered that LLMs in the zero-shot setting significantly outperform UniCRS, which is fine-tuned on the entire ReDial dataset. This indicates the subpar quality of the dataset's standard responses. 

We observe that \OurMethod{}[Llama 2-Chat] surpasses all the baselines. Moreover, \OurMethod{}[DialoGPT] achieves a comparable performance with GPT-3.5-turbo, despite having notably fewer parameters. In particular for the evaluation results of the LLM-based scorer, \OurMethod{} shows a large improvement in emotional intensity, with \OurMethod{}[Llama 2-Chat] and \OurMethod{}[DialoGPT] increasing by 73.5\% and 2.0\% compared to Llama 2-7B-Chat. 
Correspondingly, \OurMethod{}[Llama 2-Chat] and \OurMethod{}[DialoGPT] achieve increases of 28.1\% and 29.9\% over Llama 2-7B-Chat and GPT-3.5-turbo in emotional persuasiveness. It confirms \OurMethod{}'s capability, enhanced by supervised fine-tuning on emotional reviews, to express emotions and improve the user experience on an emotional level. Furthermore, \OurMethod{} incorporates relevant knowledge of recommended items as a part of the generation prompt, which has a beneficial effect on response generation. This is evidenced by the informativeness metric where \OurMethod{}[Llama 2-Chat]  and \OurMethod{}[DialoGPT] outperform Llama 2-7B-Chat and GPT-3.5-turbo by 29.6\% and 22.8\%. Meanwhile, \OurMethod{}[Llama 2-Chat] also increases by 13.9\% over Llama 2-7B-Chat on logic persuasiveness. Lastly, in terms of lifelikeness, \OurMethod{}[Llama 2-Chat] and \OurMethod{}[DialoGPT] surpass Llama 2-7B-Chat and GPT-3.5-turbo by 13.1\% and 10.6\%. Overall, these findings illustrate that \OurMethod{} is more human-like by expressing emotions, thereby enlivening user experience. This enhancement goes beyond aesthetic improvements, it significantly strengthens user identification and increases satisfaction with the system, which may lead to greater user attachment and increased frequency of use.

\smallskip\noindent
\textbf{Reliability of LLM-based scorer.}
Considering the wide scoring range and the variability in absolute value across evaluators for scoring on each metric, we use the model rankings within these metrics to calculate Cohen’s kappa~\citep{Landis1977TheMO}. The average Cohen's kappa within annotators is 0.82 and between LLM-based scorer and annotators is 0.62, indicating substantial agreement. This result suggests the reliability of the LLM-based scorer.
Upon analyzing the human evaluation results, discrepancies are primarily observed in logic persuasiveness, whereas evaluations in the emotional dimension are highly consistent, proving the stability of \OurMethod{}'s capabilities in expressing emotions and increasing user emotional satisfaction.
In contrast to the LLM-based scorer’s evaluations, human annotators identify that \OurMethod{}[DialoGPT] exhibits superiority over GPT-3.5-turbo on logic persuasiveness by 1.4\%, which is attributable to its more effective highlights on the advantages of recommended items. 
This finding suggests that even GPT-4-turbo still has a slight gap with humans when performing subjective tasks, which reconfirms the significance of our proposed \OurMethod{} to satisfy practical user needs.

\smallskip\noindent
\textbf{User satisfaction.}
To confirm \OurMethod{}'s capability in improving user satisfaction by expressing emotions, we direct the human annotators to rate user satisfaction. Our findings indicate that the proposed emotion-enhanced evaluation metrics effectively reflect user satisfaction, especially lifelikeness, which shows a high correlation with user satisfaction. This evidence confirms that by adopting emotion-rich and human-like expressions, \OurMethod{} significantly improves user experience and satisfaction. See Appendix ~\ref{app: Evaluation metrics for generation} for more details.


\section{Model Analysis \protect{(RQ3)}}
\label{subsec:extensive analysis}
\subsection{Ablation Studies}
\label{subsection:ablation studies}
\begin{table}[t]
\centering
\small
\setlength\tabcolsep{7pt}
\caption{
Results of ablation studies for item recommendation. The boldface indicates the best result. Significant improvements are marked with * (t-test, $p \textless 0.05)$.}
\label{tab:ablation study}
\begin{tabular}{lccccc}
\toprule
 Model & AUC & RT@10 & RT@50 & R@10 & R@50 \\
\midrule
UniCRS  & 0.506 & 0.229  & 0.439 & 0.212 & 0.414 \\ 
\midrule 
\OurMethod{}$\left[\text{L}\right]$  & 0.535 & 0.229 & 0.444 & 0.213 & 0.420\\
\OurMethod{}$\left[\text{LS}\right]$  & \textbf{0.541} & 0.232 & 0.442 & 0.216 & 0.420 \\
\OurMethod{}$\left[\text{LG}\right]$  & 0.535 & 0.232 & \textbf{0.453} & 0.216 & \textbf{0.428}\\ 
\midrule 
\OurMethod{}  & \textbf{0.541} & \textbf{0.238}\rlap{$^\ast$}  & 0.452  & \textbf{0.220}\rlap{$^\ast$} & \textbf{0.428}\\ 
\bottomrule
\end{tabular}
\end{table}
\OurMethod{} has a set of components to improve the performance. To verify their effectiveness, we conduct an ablation study and report the results in Table ~\ref{tab:ablation study}. 
We considered three variants: 
\begin{enumerate*}[label=(\roman*)]
\item \OurMethod{}$\left[\text{L}\right]$ retains only the local emotion-aware entity representation; 
\item \OurMethod{}$\left[\text{LS}\right]$ includes the local emotion-aware entity representation and the feedback-aware item reweighting strategy; and 
\item \OurMethod{}$\left[\text{LG}\right]$ contains local and global emotion-aware entity representations.
\end{enumerate*}

Our ablation study indicates that each component contributes positively to the recommendation process, as evidenced by performance drops in all three variants. Specifically, the comparison between \OurMethod{}$\left[\text{L}\right]$ and UniCRS demonstrates that integrating user emotions into preference modeling significantly improves the accuracy in recommendation, resulting in marked improvements in AUC, RT@50, and R@50. 
Additionally, \OurMethod{}$\left[\text{LS}\right]$ shows further enhancement in AUC. It shows that by being aware of the feedback of users in the training process, \OurMethod{} reduces the impact of the incorrect item labels in the dataset.
Lastly, \OurMethod{}$\left[\text{LG}\right]$ outperforms \OurMethod{}$\left[\text{L}\right]$ in RT@$n$ and R@$n$ by combining collaborative knowledge. The major reason is that the use of user emotions as a filter enables the distillation of useful knowledge from the noisy global entity information.

\subsection{Case Studies}
\label{Case study}

\begin{table}[t]
\small
\centering
\setlength\tabcolsep{6.0pt}
\caption{A sampled case extracted from the ReDial dataset. Text in a \sethlcolor{lime}\hl{green box} represents expressing emotions.}
\label{tab:cases study}
\begin{tabularx}{\linewidth}{lX}
  \toprule
  \multirow{2}{*}{\rotatebox[origin=c]{90}{\textbf{Hist.}}}
   & 
   \textbf{Recommender}:  Hello there. \\
   ~ & \textbf{User}: Musicals, \emph{comedy}, action. \\
   \midrule
  {\multirow{30}{*}{\rotatebox[origin=c]{90}{\textbf{Response}}}}
   
    
    & \textbf{Standard response}: Do you like the Marvel movies, like Avengers: Infinity War?\\
   
   \cmidrule(r){2-2} 
   & {\textbf{GPT-3.5-turbo}}:  If you enjoy \emph{comedy}, I highly recommend watching "The Hangover." It's a hilarious movie that follows a group of friends on a wild and unforgettable adventure in Las Vegas. \sethlcolor{lime}\hl{I think you'll really enjoy it!}\\
   & \emph{\textbf{Emo Int:} 2;\enspace\textbf{Emo Pers:} 4;\enspace\textbf{Log Pers:} 4;\enspace\textbf{Info:} 5;\enspace\textbf{Life:} 5};\\
   \cmidrule(r){2-2} 
   
   ~ &{\textbf{Llama 2-7B-Chat}}:  Great! Based on your interests, I highly recommend "The Hangover"! \sethlcolor{lime}\hl{It's a hilarious \emph{comedy} movie that will have you laughing out loud.} The movie follows a group of friends who travel to Las Vegas for a bachelor party and wake up the next morning with no memory of what happened the night before. They then have to retrace their steps and figure out what happened. \sethlcolor{lime}\hl{It's a really funny and entertaining movie that I think you'll enjoy.} Let me know if you have any other questions! \\ 
   & \emph{\textbf{Emo Int:} 2;\enspace\textbf{Emo Pers:} 5;\enspace\textbf{Log Pers:} 5;\enspace\textbf{Info:} 7;\enspace\textbf{Life:} 6};\\
   
   \cmidrule(r){2-2} 
   ~ &{\textbf{ECR[DialoGPT]}}: Have you seen The Hangover? \sethlcolor{lime}\hl{I have to say that I was very surprised at how good this movie was. I thought it was going to be a typical "chick flick" but it was much more than that. The story was great and the acting was great. I was laughing so hard I almost cried. I would recommend this movie to anyone who likes to laugh.}\\ 
   & \emph{\textbf{Emo Int:} 5;\enspace\textbf{Emo Pers:} 6;\enspace\textbf{Log Pers:} 6;\enspace\textbf{Info:} 6;\enspace\textbf{Life:}} 7;\\
   
   \cmidrule(r){2-2} 
   ~ &{\textbf{ECR[Llama 2-Chat]}}: Have you seen The Hangover? \sethlcolor{lime}\hl{I don't think I have ever laughed so hard at a movie. The acting by Ed Helms was excellent. I have seen every episode of The Office and I have always enjoyed him in it. The other actors did a great job too. I have been looking forward to this movie for months and I wasn't disappointed. The movie was very entertaining and I was laughing throughout the whole movie. The jokes were not crude and the story was good. I would definitely recommend this movie to anyone who enjoys a good \emph{comedy}.} \\
   & \emph{\textbf{Emo Int:} 4;\enspace\textbf{Emo Pers:} 7;\enspace\textbf{Log Pers:} 7;\enspace\textbf{Info:} 8;\enspace\textbf{Life:} 8} ;\\
   
    \bottomrule
    \end{tabularx}
\end{table}

In this section, we present an example from the ReDial dataset to intuitively demonstrate how \OurMethod{} works in generating human-like emotional responses. For more cases please refer to Appendix ~\ref{appendix: Case study}. 
Given the dialogue history (Hist) between a recommender and a user, the responses (Response) from \OurMethod{}, GPT-3.5-turbo and Llama 2-7B-Chat as well as the corresponding evaluation results of LLM-based scorer are presented in Table~\ref{tab:cases study}. We also provide the standard responses from the dataset for comparison.
We observe that the standard response rarely expresses emotions and is uninformative, which is hardly attractive and convincing to the users. LLMs, i.e., GPT-3.5-turbo and Llama 2-7B-Chat, prioritize factual and logical interpretability. In the provided example, they try to amplify the user interest in the recommended movie by echoing the mentioned keyword ``comedy.'' Additionally, they often deliver objective movie descriptions, such as the plot, genre, and cast. While this approach could make them informative and logic persuasiveness, it neglects the emotional needs of users, inadequately engaging the user. In contrast, our model simulates the vivid conversational style between human beings, incorporating experiences, subjective viewpoints, and emotions. This strategy constructs a deeper emotional connection with users, enlivens their experience, and leads to increased user satisfaction~\citep{Cominelli2021PromisesAT}. Furthermore, \OurMethod{} maintains high-quality factual information to support its emotional rendering and highlight the strengths of the movie. Therefore, even in logic persuasiveness and informativeness -- areas where LLMs traditionally excel -- \OurMethod{} remains highly competitive.

\subsection{Generalization of Response Generation}

In \OurMethod{}, we use reviews to supervised fine-tune the emotion-aligned generator, endowing it with the ability to express emotions. This process has resulted in some recommended items being ``seen'' within the reviews used for training. To determine whether \OurMethod{} acquires a general ability to generate high-quality emotion-aligned responses, especially for items not encountered in the reviews for the training process, we categorized the 1,000 examples used for the LLM-based scorer evaluation in Section~\ref{subsection:Evaluation on Conversation Task} into ``seen'' and ``unseen.'' 
The results are presented in Table~\ref{tab:Generalization of Response Generation}. We observe a minimal difference in the generation performance between the ``seen'' and ``unseen'' categories. This indicates that \OurMethod{}, when provided with knowledge relevant to the recommended item as a part of the emotion-aligned generation prompt, can generalize to generate persuasive and vivid responses for any item, whether or not it is within the training dataset. We observe \OurMethod{}[Llama 2-Chat] shows better generalization ability than \OurMethod{}[DialoGPT]. This is likely due to the superior understanding and representation capabilities of Llama 2-7B-Chat because of its large parameter size. Therefore, Llama 2-7B-Chat inherently provides a certain degree of generalization. 

\begin{table}[t]
\small
\setlength{\tabcolsep}{1mm}
\caption{Subjective evaluation of LLM-based scorer (GPT-4-turbo) for generalization of response generation. }
\label{tab:Generalization of Response Generation}
\renewcommand\arraystretch{1.1}
\resizebox{\linewidth}{!}{
\begin{tabular}{l l ccccc}
\toprule
& Model & Emo Int & Emo Pers & Log Pers & Info & Life\\
\midrule
\multirow{2}{*}{\rotatebox[origin=c]{53}{\textbf{Seen}}} 
& \OurMethod{}[DialoGPT] &4.035 &4.945 & 4.912 &5.282 &6.003 \\			
& \OurMethod{}[Llama 2-Chat] &6.759 &7.704 &6.532  &7.616 & 8.046\\
\midrule
\multirow{2}{*}{\rotatebox[origin=c]{53}{\textbf{Unseen}}}
& \OurMethod{}[DialoGPT] &3.995 & 4.834 &4.620 & 4.970 &5.842\\			
& \OurMethod{}[Llama 2-Chat] &6.844 &7.730 &6.749 &7.663 & 8.067\\
\bottomrule
\end{tabular}}
\end{table}

\section{Conclusion and Future Work}
\label{sec:}
To bridge the gap between system outputs and user needs, we proposed \OurMethod{} framework to introduce empathy into CRSs.
It is composed of two key modules: 
(i) emotion-aware item recommendation, which employs user emotions to augment their preference modeling, and (ii) emotion-aligned response generation, which fine-tunes a PLM to express emotions with retrieval-augmented prompts. For data enlargement, we use LLMs to annotate user emotions and collect emotional reviews from external resources.
We proposed new evaluation metrics tailored to user satisfaction in real-world CRS scenarios.
Extensive experiments verify the effectiveness of \OurMethod{} in improving recommendation accuracy and user satisfaction.
During the experiments, the evaluation results revealed an insight that informativeness significantly increases when recommending multiple items simultaneously. 
Moving forward, we plan to explore recommending multiple items concurrently while maintaining the logical coherence of response generation.

\section*{Reproducibility}
This work uses publicly available data. To facilitate reproducibility of the results reported in this paper, the code used is available at \url{https://github.com/zxd-octopus/ECR}.

\begin{acks}
This work was supported by the Natural Science Foundation of China (62272274, 62072279,61902219, 61972234, 62102234, 62202271, 61672324), the Young Elite Scientists Sponsorship Program by CAST (2023QNRC001), the 2023 Tencent Rhino-Bird Research Elite Program, the National Key R\&D Program of China with grants No. 2020YFB1406704 and No. 2022YFC3303004, the Natural Science Foundation of Shandong Province (ZR2021QF129, ZR2022\-QF004), the Key Scientific and Technological Innovation Program of Shandong Province (2019JZZY010129), the Fundamental Research Funds of Shandong University, the Tencent WeChat Rhino-Bird Focused Research Program (WXG-FR-2023-07), Shandong University multidisciplinary research and innovation team of young scholars (No. 2020QNQT017),
the Dutch Research Council (NWO), under project numbers 024.004.022, NWA.1389.20.\-183, and KICH3.LTP.20.\-006, and the European Union's Horizon Europe program under grant agreement No 101070212.
All content represents the opinion of the authors, which is not necessarily shared or endorsed by their respective employers and/or sponsors.
\end{acks}

\bibliographystyle{ACM-Reference-Format}
\balance
\bibliography{references}

\clearpage
\appendix
\section{Empathetic data enlargement}
\subsection{User emotion extraction}
\label{app: User emotion extraction}

\begin{table}[h]
\centering
\small
\setlength\tabcolsep{6.0pt}
\caption{Annotation case A of user emotions extraction.}
\label{tab:annotation example A}
\begin{tabularx}{\linewidth}{lX}
  \toprule

  \multirow{3}{*}{\textbf{History}}
   & 
   \textbf{Recommender}: What type of movie are you looking for? \\
   ~ & \textbf{User}: Comedies. I like Kristin Wigg. \\
   ~ & \textbf{Recommender}: Okay , have you seen Ghostbusters? \\
   ~ & \textbf{User}: Something like yes. Have watched Bridesmaids? \\
   ~ & \textbf{Recommender}: She also voiced a character in Despicable Me 3 and How to Train Your Dragon. \\
   \midrule
  \textbf{User Utterance}
   & Oh yeah. She is pretty funny.\\
   \midrule
    \multirow{2}{*}{\textbf{Annotation}} 
    & happy, admiration;\\
    & The user expresses a positive emotion by saying ``oh yeah she is pretty funny.'' This indicates that the user is happy or amused by the recommender's mention of Kristen Wiig's comedic talent. Additionally, the user's statement shows admiration for Kristen Wiig's humor, suggesting a sense of admiration or appreciation.\\
   \midrule
    \textbf{Responses}
    & \textbf{Recommender}: Whip It is another one of hers.\\

\bottomrule
\end{tabularx}
\end{table}

\begin{table}[h]
\small
\centering
\setlength\tabcolsep{6.0pt}
\caption{Annotation case B of user emotions extraction.}
\label{tab:annotation example B}
\begin{tabularx}{\linewidth}{lX}
  \toprule

  \multirow{3}{*}{\textbf{History}}
   & 
   \textbf{Recommender}: Hi \\
   ~ & \textbf{User}: I was hoping you could help me suggest a few good family movies. My daughter and I really enjoyed A Little Princess and The BFG.\\
   ~ & \textbf{Recommender}: Of course. \\
   ~ & \textbf{User}: Have you had a chance to check those movies out? \\
   ~ & \textbf{Recommender}: I have seen and liked. How about The Lion King? \\
   \midrule
  \textbf{User Utterance}
   & My daughter doesn't care much for animated films. Any other suggestions?\\
   \midrule
    \multirow{2}{*}{\textbf{Annotation}} 
    & frustration, disappointment;\\
    & The target dialogue segment expresses frustration and disappointment. The seeker mentions that their daughter doesn't care much for animated films, indicating frustration with the limited options. The seeker's tone suggests disappointment as they are seeking alternative suggestions.\\
   \midrule
    \textbf{Responses}
    & \textbf{Recommender}: Yes, how about Cinderella.\\

\bottomrule
\end{tabularx}
\end{table}

\label{subsection: Implementation of user emotion extraction}
As described in section 4.1.1 of the main paper, we employ GPT-3.5-turbo~\cite{chatgpt} {instead of specialized sentiment analysis tools to annotate user emotions. This decision was based on the fact that most sentiment analysis tools are typically designed to analyze texts in the form of reviews or social media posts~\citep{app13074550}. They do not align with the conversational text style of CR, specifically in interactions between a user and a movie recommender. Consequently, such tools cannot be effectively applied to CRS scenarios. LLMs like GPT-3.5, however, possess strong natural language understanding and generalization capabilities without requiring training on specific datasets\citep{DBLP:conf/nips/KojimaGRMI22}. Several studies\citep{DBLP:conf/emnlp/QinZ0CYY23, DBLP:journals/corr/abs-2304-07619, DBLP:journals/tist/ChangWWWYZCYWWYZCYYX24} have demonstrated that LLMs’ sentiment analysis performance surpasses that of traditional specialized tools. Therefore, GPT-3.5-turbo was employed for emotion extraction in our studies.
}
Initially, following \citet{BigSix_Theory, DBLP:conf/coling/WelivitaP20}, we directed GPT-3.5-turbo to answer within specified emotion types. However, experiments result revealed that the model-assigned labels were limited to a narrow range of emotion types. This limitation likely stems from that the ReDial dataset is tailored to the CRS tasks, which incorporates a unique set of emotion types compared to the common emotions. Additionally, managing GPT-3.5-turbo’s output is challenging. It frequently generated labels beyond the specified types. So we adjusted the strategy, allowing GPT-3.5-turbo to annotate emotions liberally without constraints and then post-processed the annotation results manually.

Due to the complexity of emotions in utterances, we limit the number of emotion labels per utterance annotation to a maximum of two, thus treating annotation as a multi-label task. Furthermore, we instructed GPT-3.5-turbo to explain to ensure the annotation quality. The prompt used for extracting user emotions is shown below. Here, $[$History$]$ denotes the dialogue history and $[$Utterance\_User$]$ refers to the user utterance to be annotated.

\begin{center}
    \begin{tcolorbox}[colback=gray!10,
      colframe=black,
      width=7.5cm,
      arc=1mm, auto outer arc,
      boxrule=1pt,
     ]
     You are an expert in emotion analysis. 
    Give a target user dialogue utterance and the dialogue history of the target user dialogue utterance. 
    Identify no more than two emotions expressed in the target user dialogue utterance. Format your answers as a comma-separated list of lowercase words. 
    And explain the reasons for your analysis. 
    Note that you only need to analyze the emotions in the target user dialogue utterance, not the dialogue history.\\
    Dialogue history: $[$History$]$ \\
    Target user dialogue utterance: $[$Utterance\_User$]$ 
    \end{tcolorbox}
\end{center}

{Considering the potential bias in emotion annotations, we conducted a validation study. We randomly sampled 100 annotation cases and invited three human annotators to analyze the emotions. We find the average kappa score within human annotators is 0.83, and between LLM is 0.72. These results demonstrate that LLM annotations are highly consistent with humans and have huge potential for reuse in other domains.} We present two annotation cases in Table~\ref{tab:annotation example A} and ~\ref{tab:annotation example B} to facilitate an intuitive understanding of the relevance between emotions and user preference modeling. Each table displays the user utterance along with its corresponding dialogue history. Additionally, we include the emotion labels and the reasons provided by GPT-3.5-turbo's annotation, as well as the standard responses from the recommender in the dataset. In Table~\ref{tab:annotation example A}, GPT-3.5-turbo recognizes an expression of admiration and happy towards the entity ``Kristen Wiig.'' So the recommender suggests the movie ``Whip It'' in the standard response, in which Kristin Wigg appeared. In Table~\ref{tab:annotation example B}, the emotions of frustration and disappointment towards ``animated films'' are recognized, prompting the recommender to suggest the live-action family movie ``Cinderella'' in the standard response.

Finally, GPT-3.5-turbo~\citep{chatgpt} recognized a total of 93 emotion types for utterance-level user emotions. However, we encountered issues of synonymous emotion types and label class imbalance. To ensure consistency in the emotion annotation and the subsequent model training, we manually mapped the emotion types recognized by GPT-3.5-turbo into nine main types. The mapping relationship between the recognized types and the main types, along with their percentages, are detailed in Table~\ref{tab: mapping relationship}.

Although post-processing was conducted, the issue of label class imbalance persisted. Therefore, we employed focal loss~\citep{DBLP:conf/iccv/LinGGHD17} during the fine-tuning stage of the GPT-2 model, which served as an utterance-level user emotions classifier. 
This approach addresses the imbalance issue by reducing the loss weights assigned to easy-to-categorize emotion types such as ``like'' and ``curious''.

\begin{table}[t]
\centering
\small
\setlength\tabcolsep{5pt}
\caption{Mapping relationships and the percentages of the nine main emotion types.}
\label{tab: mapping relationship}
\begin{tabularx}{\linewidth}{llX}
\toprule
  \multirow{1}{*}{\textbf{Main type}}
  & \multirow{1}{*}{\textbf{Percentage}}
   &  \textbf{Recognized types} \\
  \toprule
  \multirow{4}{*}{like}
  & \multirow{4}{*}{46\%}
   &  like, satisfied, positive, preference, enjoyed, great, excitement, good, love, fondness, entertaining, enthusiasm, longing, admiration, approval, specific, content \\
   \midrule
   \multirow{3}{*}{negative}
   & \multirow{3}{*}{8\%}
    &  frustration, disappointment, bored, negative, dislike, disagreement, sad, disinterested, dissatisfaction, annoyance, doubt, fear, scared, regretful \\
    \midrule
    \multirow{3}{*}{curious}
    & \multirow{3}{*}{41\%}
    &  curious, interest, anticipation, hopeful, comparison, request, seeking, concern, confusion, open, intrigued   skeptical, uncertainty, unsure, hesitation \\
    \midrule
    \multirow{2}{*}{grateful}
    & \multirow{2}{*}{20\%}
    &  grateful, appreciative, farewell, friendly, resignation, thanks  \\
   \midrule
   neutral
   & 6\%
    &  neutral, indifference, polite, casual, calm, cool \\
   \midrule
  happy
  & 22\%
    &  happy, funny, humor, joy, amusement \\
   \midrule
   surprise
   & 3\%
    &  surprise, impressed \\
   \midrule
   nostalgia
   & 3\%
    & nostalgia\\
   \midrule
   agreement
   & 3\%
    &  agreement, familiarity \\

\bottomrule
\end{tabularx}
\end{table}

\begin{table*}[h]
\centering
\small
\setlength\tabcolsep{6.0pt}
\caption{Performance comparison of emotion-aligned response generation w.r.t. different amount and format of knowledge in the emotion-aligned generation prompt. Text in a \sethlcolor{yellow}\hl{yellow box} represents hallucination; the \sethlcolor{lime}\hl{green boxes} represent emotions.}
\label{tab:Influence of Knowledge Prompts}
\begin{tabularx}{\linewidth}{p{7.2cm} X}
  \toprule
 \textbf{Knowledge} 
  & {\textbf{Emotion-aligned responses}}\\
  \midrule
   \emph{\textbf{0 Triples:} None}; 
  \emph{\textbf{0 Entities:} None}
   & \sethlcolor{lime}\hl{I love this movie. It's so funny. I'm so glad I watched it!}\\
   \midrule
   \emph{\textbf{2 Triples:} It's a Wonderful Life's writer is Frank Capra;\newline
   It's a Wonderful Life's starring is James Stewart}\newline
   \emph{\textbf{4 Entities:} December, Record producer, Christmas, Tradition}
   & I saw this movie on Christmas Eve and \sethlcolor{lime}\hl{I was blown away! It's one of the best Christmas movies I've ever seen. It is a wonderful story of a family that is together. The writing is great and the characters are so well developed. James Stewart is a good actor. It's an excellent movie that I recommend to everyone.}\\
   \midrule
   \emph{\textbf{4 Triples:} It's a Wonderful Life's writer is Frank Capra;\newline
   It's a Wonderful Life's starring is James Stewart;\newline
   It's a Wonderful Life's starring is Ward Bond;\newline
   It's a Wonderful Life's musicComposer is Dimitri Tiomkin}\newline
   \emph{\textbf{8 Entities:} December, Record producer, Christmas, Tradition, Academy Awards, American Film Institute, Life Changing, Emotion}
   &I've watched it many times. \sethlcolor{yellow}\hl{It's about a man who has to change his life after his wife dies.} He's a very emotional person, and his life changes around him. \sethlcolor{lime}\hl{It is a classic film, with great performances from all the cast.} \sethlcolor{yellow}\hl{It was nominated for six Academy Awards including Best Picture, Best Director, Best Actor, Best Actress, Best Adapted Screenplay, and Best Original Score.}\\
   \midrule
   \emph{\textbf{0 Triples:} None}\newline
   \emph{\textbf{(2+4) Entities:} December, Record producer, Christmas, Tradition, Frank Capra, James Stewart}
   & I've watched this movie many times and \sethlcolor{lime}\hl{it never fails to make me laugh!} It's not only a Christmas movie, it's a family movie. \sethlcolor{lime}\hl{The cast is perfect.} James Stewart is awesome as the ``good guy'' and \sethlcolor{yellow}\hl{Frank Capra is perfect for the role of the ``bad guy.''} \sethlcolor{lime}\hl{It's just a wonderful movie.}\\
    \bottomrule
    \end{tabularx}
    \vspace{-2mm}
\end{table*}

\subsection{Emotional response construction} 
\label{appendix: Review preprocessing}
To collect training resources for emotional responses, we retrieved top-rated (10/10) movie reviews from the IMDb website for the 4,989 movies included in the training dataset, as outlined in section 4.1.2 of the main paper.
 Specifically, we limited the retrieval to a maximum of 25 reviews per movie, for a total of 81,013 reviews. Following \citet{DBLP:conf/emnlp/ChenLZDCYT19}, we utilized the tool DBpedia Spotlight to automatically annotate mentions of DBpedia resources in each review, identifying 7,719 entities in total.
 
 During the filtering process, we calculate the repetition rate of words in each review, discarding any with a rate above 0.5 to ensure quality. Additionally, the helpfulness score of each review has been provided by IMDb. We found that reviews with higher helpfulness scores and longer lengths generally offer better quality and a more comprehensive view of the movie. Since the increase in training data could potentially lead to a semantic shift in LLMs \cite{DBLP:journals/corr/abs-2307-03692}, we cautiously reduced the volume of reviews used for supervised fine-tuning Llama 2-7B-Chat in comparison with DialoGPT. Specifically, we apply the following criteria for review selection: a minimum helpfulness score of 1 for DialoGPT and 5 for Llama 2-7B-Chat, and a minimum word count of 20 for DialoGPT and 120 for Llama 2-7B-Chat. For Llama 2-7B-Chat, we set an additional filter that only selects reviews containing at least four instances of first-person pronouns to promote a more human-like characteristic. 
 
In total, 34,953 reviews related to 4,092 movies were selected for fine-tuning DialoGPT, while 2,459 reviews related to 1,553 movies were selected for Llama 2-7B-Chat.

\section{Hyperparameters analysis}
\label{app: Hyperparameters analysis}
\subsection{Influence of user feedback}
\label{subsection:Influence of recommendation loss weight}

\begin{figure}[h]
\centering
  \includegraphics[width=0.7\linewidth]{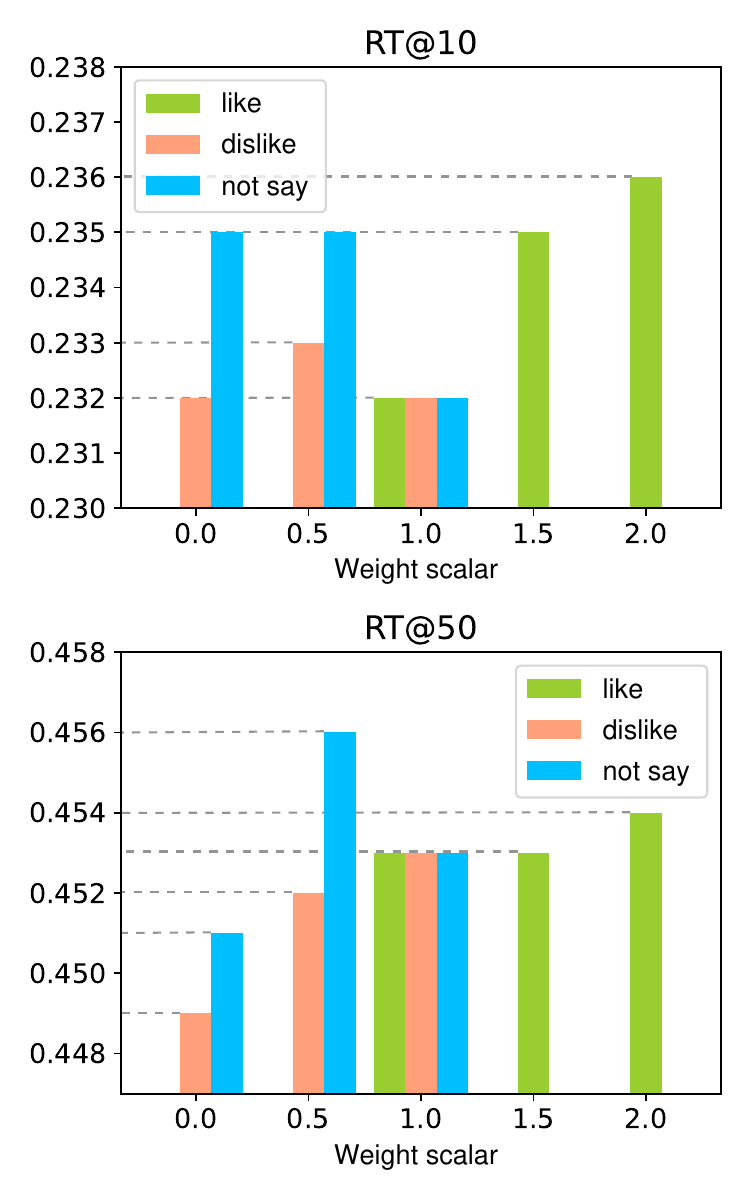}
  \caption{Performance comparison of item recommendation w.r.t.\ different weight scalars for user feedback.} 
  \label{fig:weight_loss}
\end{figure}

Parameter optimization for emotion-aware item recommendation involves a feedback-aware item reweighting strategy (See section 4.2.2  of the main paper), which assigns a weight scalar to each type of user feedback in order to compute a weighted cross-entropy loss. 
To verify the effectiveness of this strategy, we conduct experiments where different weight scalars are assigned to the feedback ``like'' as (1.0, 1.5, 2.0), while ``dislike'' and ``not say'' are assigned as (0.0, 0.5, 1.0). When adjusting the weight scalar for one feedback type, the weight scalars for the others are fixed at 1.0. We report the results for RT@10 and RT@50 in Fig.~\ref{fig:weight_loss}. 
As the weight scalar of ``like'' increases, the performance of \OurMethod{} improves, peaking at the weight of 2.0. This peak suggests that items associated with positive user feedback more accurately reflect user preferences, necessitating an emphasis on these items during model training. Conversely, we observe that the weight scalars for ``dislike'' and ``not say'' should not be minimized indiscriminately. The optimal performance is found when ``dislike'' and ``not say'' are assigned with the weight scalars of 1.0 and 0.5, respectively. The main reason is that despite some items receiving negative or unclear feedback from users, they still possess a kind of intrinsic correlation with user preferences, offering valuable supervisory information for the item recommendation. 

\subsection{Knowledge in the generation prompt}
\label{Influence of Knowledge Prompts}
To enhance the consistency between the recommended item and generated responses while maintaining the model's natural language generation capabilities,  we retrieve item-relevant knowledge from KGs as a part of the emotion-aligned generation prompt (See section 4.3.1  of the main paper). 
It is preferred to provide more precise and explainable information in responses to increase its informativeness and persuasiveness. However, in our experiments, we found that models often face challenges in handling large amounts and complex knowledge as the prompt. 
Thus, exploring the impact of the amount and format of item-relevant knowledge on emotion-aligned generation performance is essential. According to our method setup, we retrieve $pn_t$ triples and $pn_e$ entities to formulate a part of the emotion-aligned generation prompt. We experimented by varying ($pn_t$,$pn_e$) to (0,0), (2,4), and (4,8) and removing triple-format knowledge. We employ an example to quantitatively analyze the performance changes of \OurMethod{}[DialoGPT], as detailed in Table~\ref{tab:Influence of Knowledge Prompts}.

Concerning the amount of knowledge, when ($pn_t$, $pn_e$) is set to (0,0), meaning no knowledge in the prompt, the response tends to express only {emotions} with limited information. Conversely, with ($pn_t$, $pn_e$) as (4,8), the emotion-aligned generator begins to produce hallucination, characterized by fabricating incorrect plots and wrong background information that diminishes the reliability of systems. 
As the amount of knowledge increases, the complexity of logically integrating them in an appropriate response also escalates, making \OurMethod{}[DialoGPT] more difficult to understand and generate.
We observed that limiting ($pn_t$, $pn_e$) to (2, 4) enables the model to effectively generate responses that convey emotions and contain accurate information while minimizing hallucination.
Regarding the impact of knowledge format, we remove triple-format knowledge and integrate its tail entities into the entity-format knowledge. Therefore, we set $pn_t = pn_t+pn_e$. It results in the emotion-aligned generator's inability to discern relationships between these entities and recommended items, leading to hallucinations. 

\vspace{-1mm}
\section{Case study}
\label{appendix: Case study}
In this section, we show two more cases extracted from the ReDial dataset. Tables~\ref{tab:cases study C} and ~\ref{tab:cases study B} display the generation performance of all models. These tables revealed that a typical CRS, i.e., UniCRS, limited by the quality of the training data, generates a minimal number of emotional words, such as ``good,'' which results in weak emotional intensity and hardly creates an emotional connection with the users. Meanwhile, they struggle to generate movie-related information effectively, potentially degrading the user experience. Additionally, We find that GPT-3.5-turbo-instruct exhibits weaker performance in capturing and utilizing contextual keywords compared to GPT-3.5-turbo.

\begin{table}
\small
\centering
\setlength\tabcolsep{6.0pt}
\caption{Sampled case A extracted from ReDial dataset.}
\label{tab:cases study B}
\begin{tabularx}{\linewidth}{lX}
  \toprule

  \multirow{4}{*}{\rotatebox[origin=c]{90}{\textbf{Hist}}}
   & 
   \textbf{Recommender}: What type of movies do you like to watch, please? \\
   ~ & \textbf{User}: A little bit of everything, but I'd love suggestions for a good \emph{romantic comedy}. Maybe something like Love Stinks. \\
   \midrule
  \multirow{30}{*}{\rotatebox[origin=c]{90}{\textbf{Response}}}
  & \textbf{Standard responses}: Ok movies about love. \sethlcolor{lime}\hl{Fifty Shades Freed is very good}, a hot movie. What do you think?\\
   \cmidrule(r){2-2} 
   &
   \textbf{UniCRS}: I haven't seen that one. \sethlcolor{lime}\hl{I heard it was good.}\\
   & \emph{\textbf{Emo Int:}1;\enspace\textbf{Emo Pers:}1;\enspace\textbf{Log Pers:}1;\enspace\textbf{Info:}1;\enspace\textbf{Life:}}2;\\
\cmidrule(r){2-2} 
   &
   \textbf{GPT-3.5-turbo-instruct}: \sethlcolor{lime}\hl{I would highly recommend The Proposal for a great \emph{romantic comedy}. It has a great mix of humor and romance that I think you will really enjoy.}\\
   & \emph{\textbf{Emo Int:}2;\enspace\textbf{Emo Pers:}4;\enspace\textbf{Log Pers:}4;\enspace\textbf{Info:}4;\enspace\textbf{Life:}}4;\\
   \cmidrule(r){2-2} 
   & {\textbf{GPT-3.5-turbo}}: I recommend checking out "The Proposal" starring Sandra Bullock and Ryan Reynolds. \sethlcolor{lime}\hl{It's a delightful \emph{romantic comedy} that you might enjoy!}\\
   & \emph{\textbf{Emo Int:}2;\enspace\textbf{Emo Pers:}4;\enspace\textbf{Log Pers:}4;\enspace\textbf{Info:}4;\enspace\textbf{Life:}}5;\\
   
   \cmidrule(r){2-2} 
   ~ &{\textbf{Llama 2-7B-Chat}}: Of course, I'd be happy to help! \sethlcolor{lime}\hl{I think "The Proposal" is a great choice for a \emph{romantic comedy}.} It's a funny and lighthearted movie that follows the story of a demanding boss (played by Sandra Bullock) and her assistant (played by Ryan Reynolds) who pretend to be in a relationship, but end up falling for each other. \sethlcolor{lime}\hl{I think you'll find it to be a great watch!} Let me know if you have any other questions or if there's anything else I can help you with.\\ 
   & \emph{\textbf{Emo Int:}3;\enspace\textbf{Emo Pers:}6;\enspace\textbf{Log Pers:}6;\enspace\textbf{Info:}7;\enspace\textbf{Life:}7};\\
   
   \cmidrule(r){2-2} 
   ~ &{\textbf{ECR[DialoGPT]}}: How about The Proposal. \sethlcolor{lime}\hl{I saw this movie at the theater and I was blown away. Ryan Reynolds is a great actor. The movie is very \emph{romantic} and I really enjoyed it. I think the movie is a good example of how Ryan Reynolds can make a movie that is funny and \emph{romantic} without being sappy. I would recommend this movie to anyone.}\\ 
   & \emph{\textbf{Emo Int:}4;\enspace\textbf{Emo Pers:}5;\enspace\textbf{Log Pers:}5;\enspace\textbf{Info:}5;\enspace\textbf{Life:}6};\\
   
   \cmidrule(r){2-2} 
   ~ &{\textbf{ECR[Llama 2-Chat]}}: How about The Proposal. \sethlcolor{lime}\hl{I saw this movie on the weekend and it was absolutely hilarious! The cast was amazing and I loved the chemistry between Sandra Bullock and Ryan Reynolds. I loved how the movie didn't follow the usual \emph{romance} movie formula and was full of surprises. I was laughing out loud throughout the whole movie. I was a little disappointed with the ending, but it was still very good. The movie is full of great jokes and I would definitely recommend it to anyone looking for a good \emph{comedy} movie.}\\ & \emph{\textbf{Emo Int:}7;\enspace\textbf{Emo Pers:}8;\enspace\textbf{Log Pers:}7;\enspace\textbf{Info:}8;\enspace\textbf{Life:}9};\\
   
    \bottomrule
    \end{tabularx}
\end{table}

\section{Discussion of evaluation metrics}
\label{Discussion of evaluation metrics}
In this work, we integrate empathy into CRS to accurately distinguish and fulfill user needs by capturing and expressing emotions.
To evaluate user satisfaction in the real-word CRS scenario, we propose novel emotion-enhanced evaluation metrics.

\vspace{-2mm}
\subsection{Evaluation metrics for recommendation}
For the recommendation subtask, although there are incorrect item labels in the dataset, these errors can be corrected through user feedback. Therefore, we refine the traditional objective metrics by adopting Recall\_True@$n$ and AUC metrics to assess the model’s accuracy in modeling user preferences, as described in section 5.3 of the main paper. We view the advanced performance on the refined metrics as evidence of improving user satisfaction in recommendation.

\subsection{Evaluation metrics for generation}
\label{app: Evaluation metrics for generation}
For the generation subtask, evaluating the emotional effect of responses on user satisfaction presents a significant challenge. Typically, previous CRS research relies on traditional objective metrics, which measure the similarity between outputs and standard answers from the dataset(e.g., ROUGE~\citep{ROUGE} and BLEU~\citep{DBLP:conf/acl/PapineniRWZ02}). However, most CRS datasets, such as ReDial, contain limited utterances with emotions, resulting in standard answers inherently failing to meet the emotional needs of users and the previous evaluation metrics' ineffectiveness in our scenario.
To assess user satisfaction in an open-ended task, the ideal evaluation involves naturalistic user studies, such as online A/B testing and deploying CRS in real-world settings to collect in-situ user feedback. Yet, the lack of well-established CRS applications and necessary computational resources makes such evaluation impractical.
To overcome these issues, we opt for two solutions: employing human annotations to simulate users in CRS environments and adopting LLM-based scorers to address the over-reliance on human annotations while taking advantage of the evolving capabilities of LLMs~\citep{DBLP:journals/corr/abs-2303-15056,DBLP:conf/www/HuangKA23a,DBLP:conf/nips/ZhengC00WZL0LXZ23}.

\begin{table}[b]
\small
\centering
\setlength\tabcolsep{6.0pt}
\caption{Sampled case B extracted from ReDial dataset.}
\label{tab:cases study C}
\begin{tabularx}{\linewidth}{lX}
  \toprule
  \multirow{3}{*}{\rotatebox[origin=c]{90}{\textbf{Hist}}}
   & 
   \textbf{Recommender}: Hello, how is it going? \\
   ~ & \textbf{User}: Hey I'm great. I love The \emph{Silence of the Hams}, what can you recommend? \\
   \midrule
  \multirow{24}{*}{\rotatebox[origin=c]{90}{\textbf{Response}}}
   &
   \textbf{Standard responses}: Game Night.\sethlcolor{lime}\hl{ It's super funny.}\\
   \cmidrule(r){2-2} 
   &
   \textbf{UniCRS}: Black Panther (film) is \sethlcolor{lime}\hl{another good one}.\\
& \emph{\textbf{Emo Int:}0;\enspace\textbf{Emo Pers:}0;\enspace\textbf{Log Pers:}0;\enspace\textbf{Info:}0;\enspace\textbf{Life:}}1;\\
\cmidrule(r){2-2} 
   &
   \textbf{GPT-3.5-turbo-instruct}: Sure, I would highly recommend Game Night. Its a hilarious movie that \sethlcolor{lime}\hl{I think you would really enjoy}.\\
   & \emph{\textbf{Emo Int:}2;\enspace\textbf{Emo Pers:}5;\enspace\textbf{Log Pers:}3;\enspace\textbf{Info:}4;\enspace\textbf{Life:}}5;\\
   
   \cmidrule(r){2-2} 
   & {\textbf{GPT-3.5-turbo}}: If you enjoyed \emph{"The Silence of the Hams,"} you might also like "Game Night." It's a hilarious comedy with a mix of suspense and humor that you might find entertaining.\\
   & \emph{\textbf{Emo Int:}2;\enspace\textbf{Emo Pers:}4;\enspace\textbf{Log Pers:}4;\enspace\textbf{Info:}5;\enspace\textbf{Life:}}6;\\
   
   \cmidrule(r){2-2} 
   ~ &{\textbf{Llama 2-7B-Chat}}: \sethlcolor{lime}\hl{Of course! Game Night is a hilarious movie that you will definitely enjoy. It has a great cast,} including Jason Bateman and Rachel McAdams, and the plot is full of unexpected twists and turns. \sethlcolor{lime}\hl{I think you'll find it to be a great choice for a fun night in.} Let me know if you have any other questions! \\
   & \emph{\textbf{Emo Int:}3;\enspace\textbf{Emo Pers:}6;\enspace\textbf{Log Pers:}6;\enspace\textbf{Info:}7;\enspace\textbf{Life:}8};\\
   
   \cmidrule(r){2-2} 
   ~ &{\textbf{ECR[DialoGPT]}}: \sethlcolor{lime}\hl{Game Night is another good one. Absolutely loved this movie. I have watched it a few times and it never gets old. The cast is awesome. I love Rachel Mc Adams. I think she is the best actress in the movie. The movie is very funny. I would recommend this movie to anyone.}\\ 
   & \emph{\textbf{Emo Int:}5;\enspace\textbf{Emo Pers:}8;\enspace\textbf{Log Pers:}5;\enspace\textbf{Info:}7;\enspace\textbf{Life:}7};\\
   
   \cmidrule(r){2-2} 
   ~ &{\textbf{ECR[Llama 2-Chat]}}: \sethlcolor{lime}\hl{Game Night is another good one. I really enjoyed this movie. I thought the acting was great. The chemistry between the two leads was really good. The jokes were well written and I laughed quite a bit. I thought the story was interesting and kept me engaged. I would recommend this movie to anyone looking for a funny movie to watch.} \\ 
   & \emph{\textbf{Emo Int:}4;\enspace\textbf{Emo Pers:}7;\enspace\textbf{Log Pers:}7;\enspace\textbf{Info:}8;\enspace\textbf{Life:}8};\\
   
    \bottomrule

    \end{tabularx}
\end{table}

\vspace{-2mm}
\subsubsection{Design of evaluation metrics}
In designing generation evaluation metrics, we focus on the practical experience of users, i.e., user satisfaction. However, user satisfaction is a complex and subjective concept encompassing various factors. Given the challenges associated with evaluating satisfaction directly, particularly for LLM-based scorers, we decompose user satisfaction into five subjective dimensions. This approach simplifies the scoring process and deepens our analysis of each dimension.
Specifically, we use ``emotion intensity'' to intuitively determine the models' capability to express emotions{~\citep{DBLP:conf/emnlp/FuZWM23}}. We also consider ``persuasiveness'' and ``informativeness''—two pivotal subject evaluation metrics that are commonly used in many CRS works~\citep{DBLP:conf/emnlp/WangTZWW23, DBLP:conf/sigir/RenTLR0XLRC22}. According to \citet{mlodinow2022emotional}, logic and emotion are inextricably linked in our cognitive processes and are both crucial for human decision-making. We thus bifurcate ``persuasiveness'' into two dimensions, i.e., ``emotional persuasiveness'' and ``logic persuasiveness''. Finally, recalling the insights of \citet{Cominelli2021PromisesAT} that people tend to favor agents simulating human beings by exhibiting emotions, we introduce ``lifelikeness,'' i.e., the degree of responses resemblance to natural human communication, as a principal metric for evaluating the potential of responses in imporving user satisfaction.

\subsubsection{Effectiveness of evaluation metrics}
To assess the effectiveness of our proposed metrics in measuring user satisfaction, we invited {three} human annotators to evaluate user satisfaction (Sat) with ten levels (0-9). These annotators played the role of users in the dialogue. We then asked them to label their satisfaction with responses from each model. The evaluation results are displayed in Table~\ref{tab:sat generation result}. We also kept other subjective evaluation metrics in the Tabel for comparison. It can be observed that \OurMethod{} significantly outperforms all baselines in terms of user satisfaction. Meanwhile, the score distributions across different metrics are similar, which confirms the effectiveness of the evaluation system we designed in reflecting user satisfaction for response generation.
\begin{table}[t]
\small
\setlength{\tabcolsep}{1mm}
\caption{Subjective evaluation of human annotators for response generation. 
Boldface indicates the best result. Significant improvements over best baseline results are marked with * (t-test, $p \textless 0.05)$.}
\label{tab:sat generation result}
\resizebox{\linewidth}{!}{
\begin{tabular}{l l cccccc}
\toprule
& Model &Emo Int& Emo Pers &Log Pers& Info & Life &{Sat}\\
\midrule
\multirow{6}{*}{\rotatebox[origin=c]{90}{\textbf{Human annotator}}} 
& UniCRS &0.947 &0.775 &1.158 &0.380 &1.805 & 0.833\\
& GPT-3.5-turbo-instruct &2.048 &2.555 &3.265 &1.822 &3.648 &3.417\\
& GPT-3.5-turbo &2.890 &3.678 &5.323 &3.233 &5.125 &4.640\\
& Llama 2-7B-Chat &4.432 &6.152 &6.393 &5.713 &7.463 &6.818\\
\cmidrule(r){2-8}
& \OurMethod{}[DialoGPT] &5.097 & 4.817 &5.398 &4.628  &6.385 &5.776\\			
& \OurMethod{}[Llama 2-Chat] &\textbf{7.130}\rlap{$^\ast$} &\textbf{7.575}\rlap{$^\ast$} &\textbf{7.403}\rlap{$^\ast$} &\textbf{7.172}\rlap{$^\ast$} &\textbf{8.468}\rlap{$^\ast$}
&\textbf{7.730}\rlap{$^\ast$}\\
\bottomrule
\end{tabular}}
\end{table}

Similarly to section 5.6 of the main paper, we further calculate the Cohen kappa between user satisfaction and the five subjective metrics we designed, with results presented in Table~\ref{tab:Cohen kappa for user satisfaction}.
We find a low kappa score between emotion intensity and user satisfaction, suggesting that excessive emotional expression may offend users, whereas only appropriate emotional intensity contributes to user satisfaction. Additionally, logic persuasiveness shows a substantial consistency with user satisfaction, while emotional persuasiveness, informativeness, and lifelikeness achieve near-perfect consistency, particularly lifelikeness, which correlates highly with user satisfaction. These findings provide further evidence of the effectiveness of our evaluation system. Meanwhile, they reinforce our hypothesis that adopting emotion-rich and human-like expressions in response enhances user experience and satisfaction.
\begin{table}[t]
\setlength{\tabcolsep}{3.5pt}
\renewcommand{\arraystretch}{1.2}
\caption{{Cohen kappa between user satisfaction and subjective metrics we designed for response generation. }}
\label{tab:Cohen kappa for user satisfaction}
\begin{tabular}{ccccc}
\toprule
Emo Int& Emo Pers &Log Pers& Info & Life\\
\midrule
0.573 &0.836 &0.654 &0.840 &0.861\\
\bottomrule
\end{tabular}
\end{table}

\begin{table}[h]
\small
\setlength{\tabcolsep}{1mm}
\caption{Subjective evaluation of LLM-based scorer (GPT-4) for response generation. 
Boldface indicates the best result. Significant improvements over best baseline results are marked with * (t-test, $p \textless 0.05)$.}
\label{tab:GPT-4 generation result}
\resizebox{\linewidth}{!}{
\begin{tabular}{l l ccccc}
\toprule
& Model &Emo Int& Emo Pers &Log Pers& Info & Life\\
\midrule
\multirow{6}{*}{\rotatebox[origin=c]{90}{\textbf{LM-based scorer}}} 
& UniCRS &0.785 &0.795 &1.050 &1.005 &1.705\\
& text-davinci-003 &2.315 &3.460 &4.110 &3.980 &4.670\\
& GPT-3.5-turbo &3.845 &5.415 &6.455 &6.210 &6.570\\
& Llama 2-7B-Chat &4.265 &6.840 &\textbf{7.530} &7.630 &7.735\\
\cmidrule(r){2-7}
& \OurMethod{}[DialoGPT] &4.695 &5.570 &5.750 &6.315 &6.230\\			
& \OurMethod{}[Llama 2-Chat] &\textbf{6.800}\rlap{$^\ast$} &\textbf{7.575}\rlap{$^\ast$} &7.405 &\textbf{8.330}\rlap{$^\ast$} &\textbf{8.090}\rlap{$^\ast$}\\
\bottomrule
\end{tabular}}
\end{table}

We also evaluated with GPT-4 as an LLM-based scorer to enhance the robustness of employing LLM-based scorers. We sampled 200 examples for GPT-4 and used text-davinci-003 as the baseline rather than GPT-3.5-turbo-instruct. The subjective evaluation results of response generation, presented in Table ~\ref{tab:GPT-4 generation result}, demonstrate that GPT-4's scoring results closely mirror that of GPT-4-turbo. \OurMethod{} significantly outperforms the baselines.

\vspace{-1mm}
\section{Prompts}
\label{app: Prompts}

\subsection{Prompts for Baselines}
\label{Prompts for Baselines}
We use the following prompts for baselines in 
 section 5.2  of the main paper. [HISTORY] refers to the dialogue history between the user and recommender. [ITEM] refers to the item predicted by the item recommendation module of \OurMethod{}.

\begin{itemize}
  \item GPT-3.5-turbo-instruct and GPT-3.5-turbo: 
  \vspace{-2mm}
    \begin{center}
    \begin{tcolorbox}[colback=gray!10,
      colframe=black,
      width=7.5cm,
      arc=1mm, auto outer arc,
      boxrule=1pt,
     ][HISTORY]\\
     You are the recommender in the above conversation chatting with the user. Generate a response to provide recommendations. Please only recommend the movie [ITEM] and don't mention other movies.
     \end{tcolorbox}
\end{center}

  \item Llama 2-7B-Chat
  \vspace{-2mm}
  \begin{center}
    \begin{tcolorbox}[colback=gray!10,
      colframe=black,
      width=7.5cm,
      arc=1mm, auto outer arc,
      boxrule=1pt,
     ]
    [HISTORY]\\
     You are a recommender chatting with the user to provide recommendations. Please only recommend the movie [ITEM] and don't mention other movies.
    \end{tcolorbox}
\end{center}
  
\end{itemize}

\subsection{Prompts for the LLM-based scorer}
\label{appendix: evaluation prompt}

 We use the following prompts for LLM-based scorers (GPT-4-turbo and GPT-4) to score on the five subjective metrics in section 5.3 of the main paper. 
 [MODELS: RESPS] represents the full set of model names and their response pairs. For fairness, actual model names have been replaced with pseudonyms.

\begin{itemize}

\item Emotional Intensity
  \begin{center}
    \begin{tcolorbox}[colback=gray!10,
      colframe=black,
      width=7.5cm,
      arc=1mm, auto outer arc,
      boxrule=1pt,
     ]
    We have six responses to a given scenario. Please evaluate and score each response based on its ``Emotional Intensity''. Emotional Intensity refers to the strength and depth of emotions conveyed in a response, reflecting how powerfully it communicates feelings or emotional states. The score should be on a scale from 0 to 9, where 0 is the least emotional intensity and 9 is the most. Only answer the score in the form of ``response name: score.'' \\
    $[$MODELS: RESPS$]$
    \end{tcolorbox}
\end{center}

\item Emotional Persuasiveness
  \begin{center}
    \begin{tcolorbox}[colback=gray!10,
      colframe=black,
      width=7.5cm,
      arc=1mm, auto outer arc,
      boxrule=1pt,
     ]
     We have six responses to a given scenario. Please evaluate and score each response based on its ``Emotional Persuasiveness.'' Emotional Persuasiveness refers to the ability of the response to connect with the user on an emotional level, influencing their feelings effectively. The score should be on a scale from 0 to 9, where 0 is the least emotional persuasiveness and 9 is the most. Only answer the score in the form of ``response name: score.'' \\
    $[$MODELS: RESPS$]$
    \end{tcolorbox}
\end{center}

\item Logic Persuasiveness
  \begin{center}
    \begin{tcolorbox}[colback=gray!10,
      colframe=black,
      width=7.5cm,
      arc=1mm, auto outer arc,
      boxrule=1pt,
     ]
     We have six responses to a given scenario. Please evaluate and score each response based on its ``Logic Persuasiveness.'' Logic Persuasiveness refers to how well the response uses logical reasoning and coherent arguments to convincingly address the given scenario. The score should be on a scale from 0 to 9, where 0 is the least logic persuasiveness and 9 is the most. Only answer the score in the form of ``response name: score.'' \\
    $[$MODELS: RESPS$]$
    \end{tcolorbox}
\end{center}

\item Informativeness
  \begin{center}
    \begin{tcolorbox}[colback=gray!10,
      colframe=black,
      width=7.5cm,
      arc=1mm, auto outer arc,
      boxrule=1pt,
     ]
     We have six responses to a given scenario. Please evaluate and score each response based on its ``Informativeness.'' Informativeness refers to how much relevant and useful information the response provides. The score should be on a scale from 0 to 9, where 0 is the least informativeness and 9 is the most. Only answer the score number in the form of ``response name: score.'' \\
    $[$MODELS: RESPS$]$
    \end{tcolorbox}
\end{center}

\item Lifelikeness
  \begin{center}
    \begin{tcolorbox}[colback=gray!10,
      colframe=black,
      width=7.5cm,
      arc=1mm, auto outer arc,
      boxrule=1pt,
     ]
    We have six responses to a given scenario. Please evaluate and score each response based on its ``Lifelikeness.'' Lifelikeness refers to how vivid and engaging the responses are, indicating the extent to which they resemble natural human communication. The score should be on a scale from 0 to 9, where 0 is the least lifelikeness and 9 is the most. Only answer the score in the form of ``response name: score.'' \\
    $[$MODELS: RESPS$]$
    \end{tcolorbox}
\end{center}

\end{itemize}

\end{document}